\documentclass{article}

\usepackage{arxiv}

\usepackage[utf8]{inputenc} 
\usepackage[T1]{fontenc}    
\usepackage{hyperref}       
\hypersetup{
    colorlinks=true,
    linkcolor=blue,
    citecolor=blue,
    filecolor=blue,
    urlcolor=blue,
}

\usepackage{url}            

\usepackage{booktabs}       
\usepackage{float}
\usepackage{tabularx}
\usepackage{multirow}
\newcolumntype{s}{>{\hsize=.2\hsize}X}
\newcolumntype{t}{>{\hsize=.3\hsize}X}
\newcolumntype{q}{>{\hsize=1\hsize}X}
\newcolumntype{o}{>{\raggedright}X}
\usepackage{biblatex}
\usepackage[english]{babel}

\usepackage{graphicx}
\graphicspath{ {./images/} }

\usepackage[autostyle=true]{csquotes} 
\usepackage{babel}

\DeclareNameFormat{labelname:poss}{%
  \ifcase\value{uniquename}%
    \usebibmacro{name:family}
      {\namepartfamily}
      {\namepartgiven}
      {\namepartprefix}
      {\namepartsuffix}%
  \or
    \ifuseprefix
      {\usebibmacro{name:given-family}
        {\namepartfamily}
        {\namepartgiveni}
        {\namepartprefix}
        {\namepartsuffixi}}
      {\usebibmacro{name:given-family}
        {\namepartfamily}
        {\namepartgiveni}
        {\namepartprefixi}
        {\namepartsuffixi}}%
  \or
    \usebibmacro{name:given-family}
      {\namepartfamily}
      {\namepartgiven}
      {\namepartprefix}
      {\namepartsuffix}%
  \fi
  \ifnumequal{\value{listcount}}{\value{liststop}}
    {'s}
    {}%
  \usebibmacro{name:andothers}}

\newrobustcmd*{\posscitealias}{%
  \AtNextCite{%
    \DeclareNameAlias{labelname}{labelname:poss}}}

\newrobustcmd*{\posscite}{%
  \posscitealias
  \textcite}

\newrobustcmd*{\Posscite}{\bibsentence\posscite}

\newrobustcmd*{\posscites}{%
  \posscitealias
  \textcites}

\newcommand\ReferencesList{}
\newcommand\AddtoRefsList[1]{\xdef\ReferencesList{\ReferencesList,#1}}
\AtEveryCitekey{\AddtoRefsList{\thefield{entrykey}}}

\defbibenvironment{nolabelbib}
  {\list
     {}
     {\setlength{\leftmargin}{\bibhang}%
      \setlength{\itemindent}{-\leftmargin}%
      \setlength{\itemsep}{\bibitemsep}%
      \setlength{\parsep}{\bibparsep}}}
  {\endlist}
  {\item}

\usepackage{etoolbox}
\makeatletter
\patchcmd{\blx@addbackref@i}{\c@refsection}{\c@savedrefsection}{}{}

\newcounter{savedrefsection}
\newcommand\saverefsection{%
  \protected@write\@mainaux{}{\string\setcounter{savedrefsection}{\the\c@refsection}}%
}
\makeatother

\DeclareCiteCommand{\secondaryciteauthor}{}{\printnames{labelname}}{}{}
\DeclareCiteCommand{\secondaryciteyear}{}{\bibhyperref{\printdate}}{\multicitedelim}{}

\newcommand\blfootnote[1]{
    \begingroup
    \renewcommand\thefootnote{}\footnote{#1}
    \addtocounter{footnote}{-1}
    \endgroup
}

\title{Virtual Urban Field Studies: Evaluating Urban Interaction Design Using Context-Based Interface Prototypes}

\author{
 Robert Dongas *, Kazjon Grace, Samuel Gillespie, Marius Hoggenmueller \\
  Sydney School of Architecture, Design and Planning\\
  The University of Sydney\\
  NSW 2006, Australia \\
  \AND
  Martin Tomitsch \\
  TD School\\
  University of Technology Sydney\\
  NSW 2007, Australia \\
  \AND
  Stewart Worrall \\
  Australian Centre for Field Robotics\\
  The University of Sydney\\
  NSW 2006, Australia \\
}

\begin{document}
\maketitle
\begin{abstract}
In this study, we propose the use of virtual urban field studies (VUFS) through context-based interface prototypes for evaluating the interaction design of auditory interfaces. Virtual field tests use mixed-reality technologies to combine the fidelity of real-world testing with the affordability and speed of testing in the lab. In this paper, we apply this concept to rapidly test sound designs for autonomous vehicle (AV)--pedestrian interaction with a high degree of realism and fidelity. We also propose the use of psychometrically validated measures of presence in validating the verisimilitude of VUFS. Using mixed qualitative and quantitative methods, we analysed users' perceptions of presence in our VUFS prototype and the relationship to our prototype's effectiveness. We also examined the use of higher-order ambisonic spatialised audio and its impact on presence. Our results provide insights into how VUFS can be designed to facilitate presence as well as design guidelines for how this can be leveraged.
\end{abstract}


\blfootnote{* Corresponding Author: \href{robert.dongas@sydney.edu.au}{robert.dongas@sydney.edu.au}}
\blfootnote{This is a preprint version of the work, the definitive Version of Record was published in \href{https://doi.org/10.3390/mti7080082}{https://doi.org/10.3390/mti7080082}}

\section{Introduction}

Ever since computing began to emigrate from the desktop environment, testing in real-world contexts (i.e., “field” testing) has been argued as critical~\parencite{Duh2006-zo}. The more complex and dynamic the environment in which a system will be deployed, the more important it is that testing be conducted in the physical, spatial, temporal, sensory, and social surroundings of that context. This is particularly relevant in urban interactive systems: there are few environments more complex and dynamic than cities.

The challenge for any ubiquitous computing research project has always been that field testing is typically unavailable for a wide variety of economic, temporal, technological, and safety reasons. From limitations with testing medical devices with real patients, to limitations with testing AVs on real roads, to the simple limitation of taking a prototype out of the lab before it is mature enough…the reasons to avoid field testing may seem overwhelming, regardless of how critical it is known to be. In some industries, particularly those where safety regulations are paramount (such as vehicle interfaces), there has been continued development of high-fidelity physical interfaces---sit-in simulators---as part of the prototyping process~\parencite{Weir2010-kw}. However, for those products and services without the budget of a car or aeroplane manufacturer, real field testing has always been limited in both scope and frequency.

Virtual Field Studies (VFS) is an emerging area of HCI work that present a technology-mediated solution to the high cost of real field trials~\parencite{Makela2020-mx}. VFS, also known as “context-based interface prototypes”~\parencite{Hoggenmuller2021-ae} leverage mixed-reality technologies (i.e., AR, VR, etc.) to place users in an immersive, interactive virtual environment which mimics both the deployment context and interaction model of the product or service being tested. We see this as particularly appropriate for exploring emerging urban technologies, as the complex urban environment will strongly mediate user interaction in ways that cannot be studied in a usability lab. At the same time,  the urban environment also places acute technical, financial, and safety constraints on real field trials. 

 VFS for pedestrian--AV interaction~\parencite{Hoggenmueller2022-fw} in urban environments are becoming more commonplace, but to date there have been very few published papers specifically focused on audio. In this article, we present a virtual field study in the domain of sound design for pedestrian--AV interaction. Given the novelty of the method, we then develop a set of considerations for conducting such studies in urban environments, which we refer to as Virtual Urban Field Studies (VUFS). Our design considerations include a new approach to validating the capacity of VUFS to stand in for the real environment it is intended to reproduce. Specifically, our study has the following research objectives:

\vspace{12pt}
\noindent{RO$_{1}$}: \emph{To explore the efficacy, affordances and characteristics of VUFS with a sound design and audio focus.} \vspace{12pt}

\noindent{RO$_{2}$}: \emph{To investigate the impact on user experience of overlaying virtual sounds over recorded imagery in VUFS. }\vspace{12pt}

We also present a new approach to validating the efficacy of VUFS in its capacity to mirror the real-world context it is intended to represent---its verisimilitude---based on psychological studies of the notion of “presence”. This validation is intended to sit alongside the actual measures of prototype efficacy (at whatever its intended purpose is---in our case communicating vehicle intent sonically) to ensure that the virtual setup is externally valid.

\section{Related Work}

Our study draws on and contributes to human--computer interaction (HCI) research investigating the use of virtual reality (VR) for prototyping and evaluating complex interactive urban systems. We examine and expand upon claims that virtual environments offer a high degree of realism, fidelity, and sense of presence. Our study also contributes to research in interfaces between autonomous vehicles (AVs) and pedestrians. Accordingly, we review below the literature on (1) virtual environments, (2) presence, (3) prototyping and evaluation methods in HCI, and (4) AV--pedestrian interfaces.

\subsection{Virtual Environments for Developing VUFS}\label{07-sec-VEs}

Virtual environments (VEs) strive to depict digital representations of spaces and places, allowing users to engage with and become absorbed in their surroundings~\parencite{Hale_KS_Stanney_KM2014-rp}. Although these environments cannot fully replicate real-life experiences, they serve as effective platforms for simulating comparable situations by fostering a sense of immersion. In this research, we adopt the perspective of \textcite{McMahan2013-uv}, as well as \textcite{Nilsson2016-xl}, suggesting immersion can be both perceptual (i.e., based on objective capabilities of the devices used to interact with the VE) and psychological (i.e., based on subjective experiences of the user while interacting with the VE). We acknowledge the difference between the broader notion of immersion and the more distinct concept of presence, which we explore in detail in Section~\ref{07-sec-presence}.

VR technology offers users the opportunity to interact with VEs from a first-person perspective, completely enveloping them with the sensory stimuli of the digital experience. This enhances the perceptual immersion for the user, more closely resembling actual audiovisual input from the real world by correlating with their head movements.

The development of high-quality, computer-generated VEs, especially for use in VR, is not without its challenges. The creation of such environments often demands considerable investment in terms of time, financial resources, and expertise~\parencite{Jones2022-qn}. Researchers and developers must consider the potential benefits and drawbacks of employing VEs in their specific context, ensuring that they balance the associated costs and complexities. By carefully selecting the sources of visual and audio stimuli for virtual environments, some of the investment burden can be reduced, allowing for more efficient content creation without compromising the overall quality and user experience. We discuss several specific VE technologies relevant to our study below: 360-degree video and spatial audio.

\subsubsection{360-Degree Video}

One method of creating VEs for VR is through the use of 360-degree video. This approach involves recording video content and spherically projecting it in the virtual space. Users are able to observe the entire environment through three degrees of freedom, determined by their head rotation. Prior research has highlighted the immersive qualities of 360-degree video in areas such as storytelling~\parencite{Skola2020-jk}, education~\parencite{Blair2021-mp}, and health/wellbeing~\parencite{Waller2021-pd}. These studies point to the inherent realism of the medium, which can be attributed to its ability to capture real-world locations and situations through video recordings. A major advantage of 360-degree video is that there is no lengthy asset design, development and integration process, as there is with more-traditional 3D modelled VEs~\parencite{Jones2022-qn}. 

A notable limitation of 360-degree video is the restricted perspective, as users are confined to the vantage point of the camera and cannot actively navigate or move around the VE. To prevent motion-induced sickness, it is generally advised not to alter the camera position during recording~\parencite{Patrao2015-ml}. Despite this constraint, 360-degree video remains a valuable medium, particularly in scenarios where the user is stationary and the primary focus is more on observation rather than direct interaction. Our use of this approach balances the immersive qualities of VR and the practical considerations of realistic content generation, enabling faster and more cost-effective production.

\subsubsection{Spatial Audio and Soundscapes}\label{07-sec-soundscapes}

The sense of hearing plays a crucial role in human perception, providing context and situational awareness beyond the scope of our vision. Spatial audio, characterised by its ability to convey the direction and distance of sound sources around a listener, has been utilised in various mediums such as film, television, video games, and exhibitions. In VR specifically, it can elevate users' sense of perceptual and psychological immersion within a virtual environment, especially when compared to non-spatial audio~\parencite{Poeschl-Guenther2013-em, Potter2022-fa}. 

Spatial audio can be leveraged in VR using a variety of methods. One such method is ambisonics, a technique for capturing and reproducing sound fields in three dimensions~\parencite{Fellgett1974-sl}. This allows physical acoustic properties of the sound field (e.g., directionality and pressure) to be encoded to varying orders of spatial resolution. While first-order (i.e., lower-quality approximate) ambisonics systems have seen renewed interest in recent years, Zotter~\parencite{Zotter2019-ub} argues that the advantages in interactivity offered by higher-order ambisonics as well as the limitations of first-order recording techniques are both compelling reasons to increase the directional resolution of recordings. These higher-order systems are now possible due to improvements in microphone technologies and signal processing methods~\parencite{Bertet2006-kn}, although this comes with added cost and complexity. Ambisonics offer a flexible platform for designing and producing immersive auditory soundscapes for use in VEs. \textcite{Hong2017-gw} highlights the potential of spatial audio for creating immersive and ecologically valid soundscapes in VR, which we intend to explore further in this research. One of the objectives of this study is determining what, if any, advantage in the subjective user experience is conveyed by the incorporation of ambisonic spatial audio technologies in VR experiences.

\subsection{Presence}\label{07-sec-presence}

At its most simple, any prototype designed for a virtual field study should be judged by its ability to make participants feel as if they really are in the “field” in question. Throughout the literature, researchers have presented multiple explanations of this notion of “being there” in order to conceptualise the complex psychological construct usually referred to as “presence”~\parencite{Steuer1992-at, Slater1993-mf, Biocca1997-fx, Lombard1997-lq, Ijsselsteijn2000-rf, Riva2004-wm}. Of particular relevance to this study, \textcite{Slater2009-iz} defines the terms place illusion (“the strong illusion of being in a place in spite of the sure knowledge that you are not there”) and plausibility illusion (“the illusion that what is apparently happening is really happening”) as key components of presence.

\textcite{Witmer1998-fz} posit that presence occurs when users do not notice the artificiality of objects in a simulated environment. Expanding upon this is Lee’s theory of presence, which defines it as “a psychological state in which virtual objects are experienced as actual objects in either sensory or nonsensory ways”~\parencite{Lee2004-qg}. Lee promotes a way of categorising the virtual experience by making the distinction between artificiality and para-authenticity. Objects with no real-world counterparts represent artificiality, while para-authenticity describes the extension of actual objects in a virtual environment. We agree with \posscite{Gilbert2016-wf} notion that “authenticity is the human-based factor that influences presence, as measured by whether it aligns with the expectations of users”. Lee’s model also breaks presence down into three distinct subdomains: physical (presence arising from the environment), social (presence arising from others), and self (presence arising from the body)~\parencite{Lee2004-qg}. We find this to be a suitable contextual framing, particularly within the domain of interaction design in urban settings, where there is interplay between all three~subdomains.

\subsubsection{Measuring Presence}

Although the immersiveness of VEs has been widely promulgated, translating these qualities into quantifiable and operational measures can be difficult. Evaluating the effectiveness of these environments presents a unique challenge, especially when utilising the technological medium of VR. We propose elevating the psychological measure of presence as a key quality in designing and developing VUFS (as we argue in this paper), as it has been well-operationalised in that literature and is a strong driver of immersion. These measures of presence can be applied through a variety of psychometrically validated instruments, which we apply to VE users as a way of measuring the efficacy of those VEs at recreating the desired environment.

We adopt the definition of presence as “the extent to which something (environment, person, object, or any other stimulus) appears to exist in the same physical world as the observer”~\parencite{Felton2022-np}. Understanding how to effectively measure presence according to this definition is crucial in assessing the success and validity of VEs. Presence as a subjective psychological construct requires the user to self-report their experience~\parencite{Insko2003-zh}. These data are commonly collected through post-experience questionnaires, which are low-cost and easy to administer. They also “embody self-reports and therefore gather the participants’ subjective experiences”~\parencite{Putze2020-il}. When utilising this method, the reliability and validity of the questionnaires need to be carefully designed for consistent research to avoid potential biases such as the “design of the question, questionnaire design, and administration”~\parencite{Choi2005-xs}. 
 
The Multimodal Presence Scale (MPS) encompasses each of Lee's presence constructs: physical, social, and self~\parencite{Makransky2017-kw}. It has been validated in the context of VR learning simulations through confirmatory factor analysis and item response theory. The ability of the MPS to effectively measure attributes of presence such as realism, attention and mediation strengthens its suitability for our purposes. These attributes are good indicators of involvement in the virtual experience, but also help to establish the ecological validity of the environments for VUFS. For these reasons, we adopt the MPS for use in our study.

\textcite{Souza2021-hk} provides a review of the various measurements of presence, categorising them into objective and subjective methods---the latter being mostly the questionnaire-based methods discussed above. Objective methods are “captured through data collection or observation, and can be divided into physiological and behavioural measures”~\parencite{Souza2021-hk}. Examples include measuring externally observed responses, heart rate, eye movements and skin temperature~\parencite{Hale_KS_Stanney_KM2014-rp}. While these objective measures can be said to provide data that are more reliable than self-reporting from a participant bias standpoint, the literature is not clear on whether they are more accurate as standalone measures of the sense of presence~\parencite{Grassini2020-wu}. While these physiological tools are an area worth watching for VUFS evaluation, we administer the more well-established survey instruments in this work.

\subsection{Prototyping and Evaluation Methods in Human Computer Interaction}\label{07-sec-prototypes}

The creation of prototypes is an important activity in any interaction design process~\parencite{Buxton2007-bw}, as it allows designers to explore and refine ideas before the envisioned product or service is developed in its final manifestation. Prototypes can fulfil different purposes, such as enabling designers to traverse a design space in a more generative manner, and evaluating manifestations of design ideas with stakeholders~\parencite{Lim2008-fl}. For screen-based interfaces, at an early stage of the design process, designers typically start with low-fidelity prototypes (e.g., sketches, paper prototype) to validate concepts, before moving on to more high-fidelity prototypes (e.g., realistic mockups, interactive click-throughs).

Over the past two decades, the use of interactive products and services has become increasingly interwoven in people’s everyday lives. As a consequence, designers have had to adapt how they prototype in order to consider interactions in context, in addition to traditional task-based measures of usability. \textcite{Buchenau2000-cu} coined the term “experience prototype”, which they define as “any kind of representation, in any medium, that is designed to understand, explore or communicate what it might be like to engage with the product, space or system we are designing”. In a similar vein, related terms such as “context-based interface prototyping” have been used to describe approaches of prototyping interfaces and interactions in the context that they are expected to be used~\parencite{Flohr2022-hz, Hoggenmuller2021-ae}. This incorporates a plethora of prototyping and design techniques, including bodystorming, enactments, physical prototyping, and more. 

With this turn towards the contextual in HCI, researchers have stressed the importance of evaluating prototypes “in the wild” rather than relying entirely on tests in laboratory environments~\parencite{Chamberlain2012-vb}. This has been identified as particularly relevant in the design of urban technologies~\parencite{Paulos2005-em, Hoggenmuller2014-dd}. However, conducting field studies is expensive and time-consuming, since prototypes have to be safe, stable, movable, and robust. In the context of prototyping and evaluating interfaces for autonomous vehicles, conducting user studies in the real-world could even cause harm to participants~\parencite{Nguyen2019-sn}. 

To reduce costs and risks, HCI researchers increasingly turned to video- and simulation-based approaches for evaluating prototypes through contextualised setups, also referred to as virtual field studies~\parencite{Makela2020-mx}. This includes video recordings which are displayed on a conventional screen~\parencite{Hoggenmuller2021-ae}, as well as computer-generated simulations presented to participants through projection-based VR environments (also known as CAVE)~\parencite{Sutcliffe2005-lq, Sutcliffe2018-ja} or VR headsets~\parencite{Flohr2020-vq}. Several studies across different contexts (e.g., public displays~\parencite{Makela2020-mx}, wearable technology~\parencite{Tran2023-qv} and smart home devices~\parencite{Voit2019-yp} have demonstrated that the evaluation of interfaces in immersive VR environments holds comparable usability results and user behaviour to that of real-world settings. Researchers have further found that the use of 360-degree video recordings of real-world prototypes can further improve perceived representational fidelity and sense of presence~\parencite{Yeo2020-zm}. In the context of AV--pedestrian interfaces, Hoggenmueller et~al.~\parencite{Hoggenmuller2021-ae} have shown that the use of real-world representations in VR is highly effective for the evaluation of contextual aspects and to assess overall trust. We propose that these advantages of higher visual fidelity means of prototyping suggest that a similar effect will be observable for technologies offering higher aural fidelity (i.e., ambisonic recordings). However, to the best of our knowledge, previous research has not established the efficacy of ambisonics in the context of VR prototypes. 

\subsection{AV--Pedestrian Interfaces}\label{07-sec-Interfaces}

With the uptake of autonomous vehicles expected to continue, research has increasingly turned to addressing human factor challenges in order to increase public acceptance and trust towards AV technologies~\parencite{Kaur2018-mj}. One such challenge is the lack of interpersonal communication cues (e.g., hand gestures, facial expressions, and eye contact) between AVs and other road users. Researchers have therefore proposed the use of external human--machine interfaces (eHMIs) to augment automated driving~\parencite{Dey2020-fx}. For instance, eHMIs can be used to communicate the AV’s intent to yield when negotiating right-of-way with pedestrians in street crossing scenarios~\parencite{Hollander2019-po}, or to signal the AV’s intent and awareness in shared spaces where they operate in close proximity to other road users~\parencite{Hoggenmueller2022-fw}. 

There now exist a wide range of eHMI design concepts, a majority of which use a visual communication modality~\parencite{Dey2020-fx}. These range from projections on the road~\parencite{Nguyen2019-sn} to displays attached to the vehicle itself, including light band bars~\parencite{Hoggenmueller2022-fw, Dey2020-fx} as well as screens capable of displaying text or symbols~\parencite{Hollander2019-po}. VR simulation studies have shown that eHMIs can significantly reduce the risk of collisions~\parencite{Faas2020-gx, M_Faas2021-ox} and increase pedestrians' subjective feeling of safety~\parencite{Hollander2019-po}. However, there is also increasing evidence that pedestrians mainly rely on implicit cues~\parencite{Risto2017-gr, Dey2017-tz, Moore2019-ul}, such as a vehicle’s movement, when deciding whether to cross in front of it. Consequently, sound has been suggested as an implicit communication modality, with the latest regulation frameworks requiring engine-like noise emissions for electric and hybrid cars to protect vulnerable road users~\parencite{Pelikan2023-tb}. For example, \textcite{Moore2020-be} found that adding synthetic engine sound to a hybrid autonomous vehicle led to increased interaction quality and clarity around the vehicle’s intent to yield. Pelikan et~al.~\parencite{Pelikan2023-tb} adopted a research-through-design approach to explore sound designs for autonomous shuttle buses in close-proximity interactions with other road users. Using voice-overs to video recordings of the buses, followed by Wizard of Oz testing in the real world, they discovered that prolonged jingles drew attention to the bus and encouraged interaction, whereas repeated short beeps and bell noises could be used to direct people away from the bus. At the same time, they emphasised the impact of situational and sequential contexts (e.g., the movements of the bus and other people) in determining the interaction that a specific sound may accomplish.

\section{Methods}

We combined research-through-design, an approach which establishes design knowledge through the generation and evaluation of artifacts~\parencite{Zimmerman2007-cc}, with a summative mixed-methods user study of the efficacy of our VUFS. To address RO1, we developed a context-based interface prototype, iteratively evaluating and improving it using a real-world use-case. Our final user study provided results and analysis of how effectively our virtual sounds preserved participants sense of presence in our VUFS (addressing RO2). We integrate this analysis with the insights from our design process~\parencite{Keyson2009-mx} to develop a set of considerations for future VUFS. 

In this section we describe both the design of our prototype VUFS (see Section~\ref{07-methods-des}) and the methodology of our summative user study (see Section~\ref{07-methods-study}). In the former, we detail technical specifications, design decisions, and where and what we depicted in our virtual environment. In the latter section we describe the participant demographics, study procedures and data analysis methods.

\subsection{Designing Our AV--Pedestrian Context-Based Interface Prototype}\label{07-methods-des} 

We designed and developed a prototype to evaluate auditory cues for AV--pedestrian interaction. The prototype utilised recorded 360-degree video and ambisonic audio to create a VE to simulate urban shared spaces. We chose to focus on an area with minimal segregation between transport modes, i.e., concurrent walking, cycling and driving, a known challenge for AVs~\parencite{Wang2022-zb}. Users engaged with this VE through a VR headset, adopting the perspective of the pedestrian. 

\subsubsection{Virtual AV--Pedestrian Scenarios} 

Our prototype presented a variety of scenarios in which an AV approached a stationary pedestrian, depicting potential encounters in a shared space. In each scenario, the vehicle performed one of three possible behaviours: avoiding the pedestrian by changing path; slowing down to a gradual stop in front of the pedestrian; or continuing at the same speed and theoretically colliding with the pedestrian. The scenarios were recorded with a real vehicle, autonomously performing the avoid and stop behaviours listed above. For the collision behaviour, the vehicle was operated manually, maintaining a constant speed until emergency braking as close as possible to the camera. The selected scenarios and behaviours in the study were designed to represent realistic potential AV--pedestrian interactions, particularly in a shared walking/driving/cycling multi-use space.

\begin{figure}[H]

\centering 
\includegraphics[width={\textwidth}]{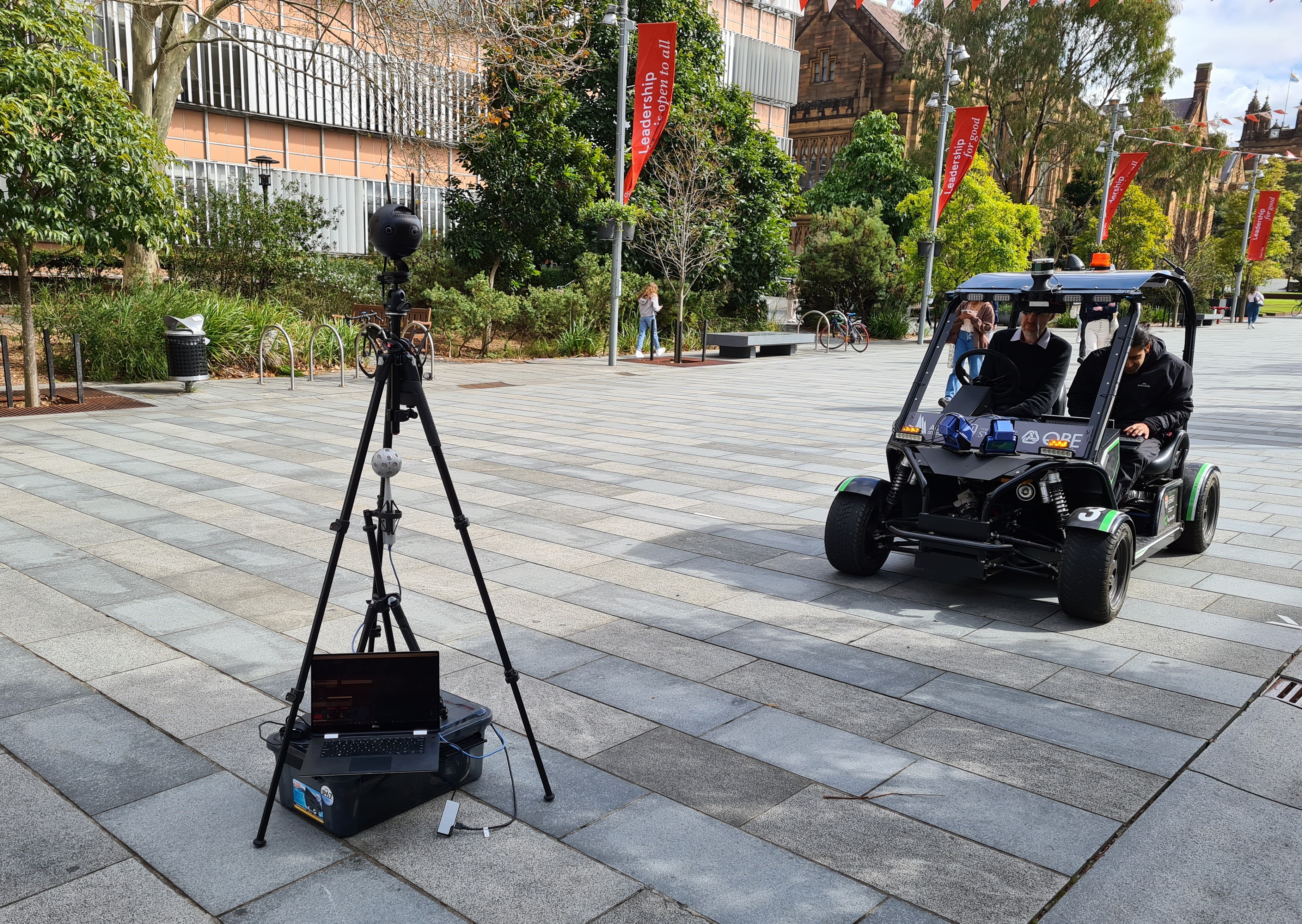}

\caption{The autonomous vehicle during technical testing, having navigated through a shared space and come to a stop. The scenario is being recorded by a 360-degree video camera and spatial microphone to create an immersive virtual environment for a Virtual Urban Field Study.\label{07-fig1}}
\end{figure}

To record the scenarios for our prototype, we used a fully functional research AV, developed by the Intelligent Transport Systems Group from the Australian Centre for Field Robotics~\parencite{Olaverri2019-ic}. The small vehicle (see Figure~\ref{07-fig1}) is specifically engineered to operate in urban environments at low speeds, aiming to manoeuvre safely through pedestrians using advanced sensors and control systems.

The avoid and stop behaviours implemented in the prototype represent the default programming of the AV depending on its mode of operation. These behaviours are designed to prioritise pedestrian safety and comfort, placing the responsibility of movement and adaptability on the vehicle. The collision behaviour was created for the purpose of researching the impact of sound design in such a scenario. Although it is (obviously) not a default behaviour of the vehicle, its inclusion in the study acknowledges the possibility that an AV may become aware of a potential collision with a pedestrian, but be unable to react in time to avoid it. Our choice of scenarios and behaviours is informed by our design to align the study with real-world urban situations, including both normal operation of an AV as well as potential exceptional cases.

\subsubsection{Virtual AV Sound Design}

The soundscape design for our prototype followed a hybrid approach \parencite[c.f.][]{Johansen2022-zg}, integrating recorded ambient environmental sounds with designed and produced vehicle sound effects. This combines the conceptualisations of soundscapes as ``perceptual constructs that emerge when a listener hears sounds from a physical environment'' and as ``a collection of sounds that are organised in an intentional way and played to listeners''~\parencite{Johansen2022-zg}. In other words, sounds are concurrently both sensory and socio-cultural experiences. A hybrid approach such as this is extremely relevant for conducting VUFS, particularly for auditory interfaces, given the need for sound composition as an intervention within the rich acoustic context of an urban environment. Our ambient recordings captured the live environmental audio for each video of each scenario that made up the virtual experience. This included many different sound sources: wind, bird song, people walking and talking, aeroplane noises, distant traffic sounds and consequential sounds of the AV (e.g., tyres on the ground). We employed fourth-order ambisonic recording techniques to capture a high-resolution spatial representation of the sound field (combining 32 discrete directional recording channels). This was mapped into a headphone mix for the VR experience, which (when combined with head tracking) enables a highly immersive audio field.

For each of the three possible AV behaviours (avoiding, stopping and colliding), three distinct sound variations were designed (S1, S2, and S3). These sound designs differed in characteristics, ranging from simpler to more complex. The stop sound aimed to convey the vehicle’s intent to decelerate to a halt without requiring any pedestrian response, while the avoid sound was designed to indicate the AV’s intent to manoeuvre around the stationary pedestrian. The collide sound, on the other hand, was intended to warn pedestrians of an impending collision, requiring them to act. Various sound characteristics (including timbre, pitch, and tempo) were manipulated to design the desired effects for each behaviour (see Table~\ref{07-tab1}). For instance, the stop sound exhibited decreasing intensity, pitch, and tempo, whereas the collide sound portrayed increasing intensity, pitch, and tempo to elicit a sense of urgency in pedestrians.

\begin{table}[H] \small
\centering
\caption{Overview of sound variations (S1, S2 and S3) designed for each of the three possible autonomous vehicle behaviours: avoiding, stopping and colliding. The sound designs exhibit distinct characteristics, with various attributes such as timbre, pitch and tempo manipulated to communicate each specific behaviour.\label{07-tab1}}

   \setlength{\tabcolsep}{6.6mm}
\begin{tabular}{lllll}
\toprule
\textbf{Behaviour}	& \textbf{Characteristic}	& \textbf{S1} & \textbf{S2} & \textbf{S3}\\
\midrule
Avoid              & Notes in pattern           & One         & Two & Three \\
\midrule
Stop               & Falling pitch              & No          & Yes & Yes   \\
                   & Diminishing tempo          & Yes         & No  & Yes   \\
                   & Reducing timbrel intensity & Yes         & Yes & Yes   \\
\midrule
Collide            & Rise in pitch              & No          & Yes & Yes   \\
                   & Increasing tempo           & Yes         & No  & Yes   \\
                   & Intensifying Timbre        & Yes         & Yes & Yes   \\
\bottomrule
\end{tabular}
\end{table}

Our sound production approach focused on applying just enough processing to situate the sound effectively in the VE. We employed some reverb to introduce approximations of diffused reflections based on suggestions by~\textcite{Jain2021-ab}, which decreased as the autonomous vehicle approached the user. More accurate simulation of the acoustic reverberation conditions would be possible through measuring multiple spatial impulse responses in the testing environment~\parencite{Murphy2014-vk}. However, given the additional complexity this would add to the audio recording, processing and implementation, we opted against this higher-fidelity processing. We also implemented an equalisation (EQ) shift and used compression to minimise the variation in dynamic range, ensuring a consistent audio experience throughout the virtual environment. Our goal here was to make the (introduced) vehicle sound effects parallel the sonic environment of the ambisonic recording as much as feasibly possible.

\subsubsection{Virtual Environment Implementation}

\begin{figure}[H]
\centering
\includegraphics[width=12cm]{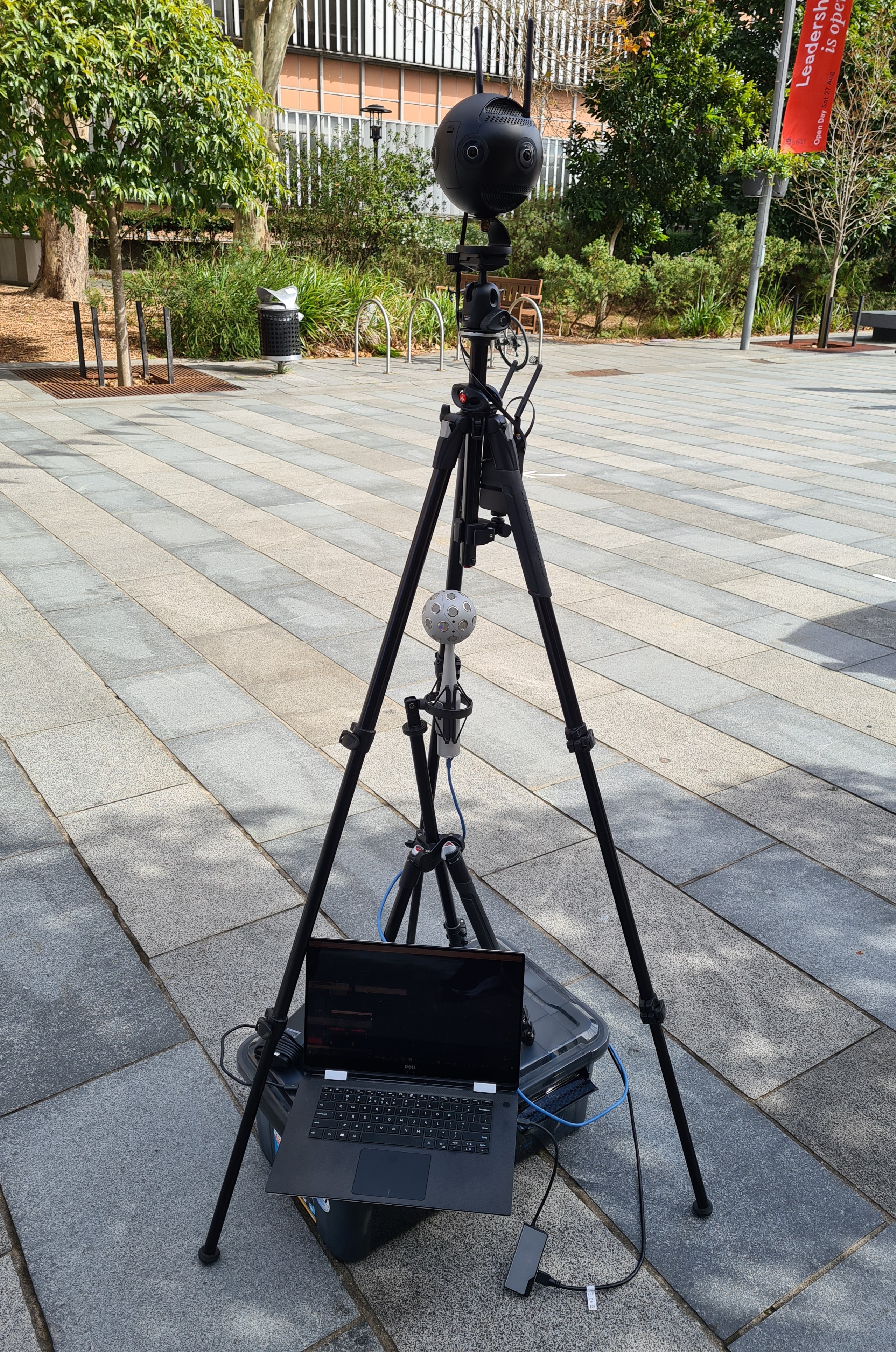}
\caption{The camera and microphone setup used for capturing 360-degree video and spatial audio, illustrating the equipment configuration and arrangement for creating immersive virtual environments used in Virtual Urban Field Studies.\label{07-fig2}}
\end{figure}

360-degree videos were recorded using an Insta360 Pro 2 camera (Maufacturer: Insta360, Shenzhen, China) set to a high frame rate (120~fps) in stereoscopic mode for capturing high fidelity images (3840~$\times$~3840 per eye). For ambient audio recording, an Eigenmike EM-32 (32-channel spherical microphone array -- Manufacturer: mh acoustics, New Jersey, United States) was utilised to capture and encode 4th-order ambisonic output. Figure~\ref{07-fig2} shows the configuration of the recording devices in the real-world location. The filming took place on a well-lit sunny day, which provided ample natural light to enhance clarity and details. The recording equipment was strategically placed within a shadowed area so its own shadow was not visible and could not distract users or detract from immersion. To mitigate microphone noise, we also recorded during periods of minimal wind. Unreal Engine, a versatile game development platform, was used to develop the virtual environment for interaction in VR by combining the 360-degree video recordings, ambient audio recordings, and overlaid vehicle sound effects.

\subsection{User Study Design}\label{07-methods-study}

\subsubsection{Participant Demographics}
We conducted a between-subjects user study of 48 participants, with 12 users in a control group and 36 users in the experimental group. The control group experienced the virtual scenarios with no overlaid vehicle sound effects, divided into two conditions of spatial (\textit{n}~=~6) and non-spatial (\textit{n}~=~6) ambient audio recordings. The control group was used to assess if the vehicle behaviours were predictable from the visuals alone, which would have confounded the evaluation of the sound’s ability to communicate vehicle intent. The experimental group completed the experience with overlaid vehicle sounds and were similarly divided into two conditions for spatial (\textit{n}~=~18) and non-spatial (\textit{n}~=~18) audio recordings.

We recruited participants by distributing flyers and advertisements around a university campus, targeting a broad spectrum of potential participants. The demographics of the participants covered a wide age range, from 19 to 53 (M = 26.8, SD = 7.5), thus providing varied perspectives in evaluating the AV--pedestrian interactions. The participants also represented diverse cultural and ethnic backgrounds, although we did not collect specific demographics in this respect. Due to recruitment limitations, our age ranges were skewed towards younger adult participants (<=25: 54\%; 26--40: 38\%; >=41: 8\%), which limits the generalisability of this study to older adults. The majority of participants (over 90\%) were familiar with the environment presented in the virtual experience, as they were staff or students at the university.

\subsubsection{Study Procedures}
Prior to each session, we informed participants of the study aims relating to the sound design of AV--pedestrian interactions, and secured their consent to participate in the study. The experience was deployed on a Vive Pro 2 headset (Manufacturer: HTC, New Taipei, Taiwan), which was adjusted and calibrated for each participant (including volume levels in the integrated headphones to account for individual hearing ability). Participants were given a short introduction on how to interact with the system using the controllers, with specific instructions to stop and remove the headset if they felt any sickness or discomfort during the study. All participants completed the full VR experience (which totalled about 15 min) without reporting any sickness or~discomfort.

The experience consisted of nine total scenarios, with three unique variations of each of the three possible AV behaviours. Given that there were nine scenarios, 162 participants would be required for a balanced Latin square design, which was a prohibitively high number. We instead paired spatial- and non-spatial-condition participants and assigned each pair a random order, counterbalancing between conditions but not fully controlling for all ordering effects. While we only sampled a subset of the orderings, the control condition results show that it was not possible to predict vehicle intent from just the video recording, which means that any learning effects could only have come from the sounds. 

\begin{figure}[H]

\centering
\includegraphics[width={\textwidth}]{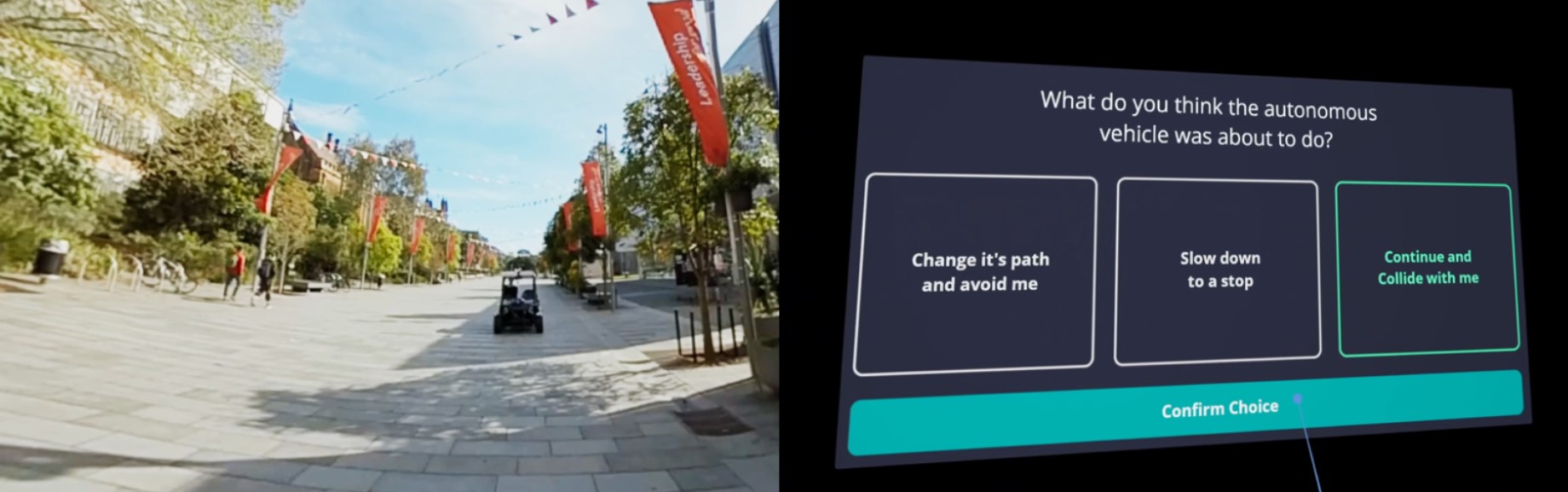}
\caption{Screenshot of the 360-degree video recording from a user's perspective in virtual reality (\textbf{left}). The panel interface enabling a user to select their prediction for the behaviour of the autonomous vehicle (\textbf{right}).\label{07-fig3}}
\end{figure}

The participants watched each 360-degree video play for approximately 20--30~s, to understand the context of the scenario as the vehicle drove towards them. The videos were paused before the behaviours of the vehicle became obvious (either through changes in speed or direction). In the experimental condition, the designed AV sounds were played at the appropriate time, synchronised to end just as the video paused. At this point an interface appeared asking the participant what they expected the vehicle behaviour to be: “Change it’s path and avoid me”; “Slow down to a stop”; or “Continue and collide with me”. The interface then displayed the question “Why did you think the vehicle was going to [chosen behaviour]?”, prompting participants to answer aloud. This flow was repeated for each of the nine scenarios. Figure~\ref{07-fig3} shows an example of what the user would see in the virtual experience.

Both quantitative and qualitative measures were employed to assess user responses and evaluate the study outcomes. Following the scenarios, participants were asked to complete a short questionnaire in VR to evaluate their sense of presence throughout the experience. The Multimodal Presence Scale~\parencite[MPS,][]{Makransky2017-kw} was used based on its validated capacity to measure users’ subjective feelings of presence within virtual environments. Evaluating presence is important to validate that our prototype was effectively leveraging the immersive qualities of the virtual environment, eliciting realistic responses from users when interacting with the AV--pedestrian scenarios. Two additional audio-specific items were added, adapted from~\posscite{Witmer1998-fz} Presence Questionnaire (PQ), as the MPS lacks any audio-specific questions. Table~\ref{07-tab2} shows the full list of questions. These questions aim to specifically investigate the influence of audio on user experience and overall immersion within the virtual environment.

\newcolumntype{i}{>{\hsize=.05\hsize}X}

\begin{table}[H] \footnotesize 
\caption{Statements from the Multimodal Presence Scale \parencite{Makransky2017-kw}, organised by subscales for physical, social and self presence. Additional audio-specific items adapted from \posscite{Witmer1998-fz} Presence Questionnaire. These statements were used in a questionnaire to evaluate users' sense of presence throughout the Virtual Urban Field Study.\label{07-tab2}}
\centering\makebox[{\textwidth}][c]{%
\begin{tabularx}{\textwidth}[t]{isX}
\toprule
\textbf{Item}	& \textbf{Source (Subscale)}	& \textbf{Statement}\\
\midrule
1       & MPS (Physical)      & The virtual environment seemed real to me.                                                      \\
2       & MPS (Physical)      & I had a sense of acting in the virtual environment, rather than operating something from outside.                          \\
3       & MPS (Physical)      & My experience in the virtual environment seemed consistent with my experiences in the real world.                          \\
4       & MPS (Physical)      & While I was in the virtual environment, I had a sense of being there.                                        \\
5       & MPS (Physical)      & I was completely captivated by the virtual world.                                                  \\
\midrule
6       & MPS (Social)       & I felt like I was in the presence of another person in the virtual environment.                                   \\
7       & MPS (Social)       & I felt that the people in the virtual environment were aware of my presence.                                     \\
8       & MPS (Social)       & The people in the virtual environment appeared to be sentient (conscious and alive) to me.                              \\
9       & MPS (Social)       & During the simulation there were times where the computer interface seemed to disappear, and I felt like I was working directly with another person. \\
10      & MPS (Social)       & I had a sense that I was interacting with other people in the virtual environment, rather than a computer simulation.                \\
\midrule
11      & MPS (Self)        & I felt like my virtual embodiment was an extension of my real body within the virtual environment.                          \\
12      & MPS (Self)        & When something happened to my virtual embodiment, it felt like it was happening to my real body.                           \\
13      & MPS (Self)        & I felt like my real arm was projected into the virtual environment through my virtual embodiment.                          \\
14      & MPS (Self)        & I felt like my real hand was inside of the virtual environment.                                           \\
15      & MPS (Self)        & During the simulation, I felt like my virtual embodiment and my real body became one and the same.                          \\
\midrule
16      & PQ (Audio~\textsuperscript{1})            & I was able to identify different sounds and distinguish them from one another.                                    \\
17      & PQ (Audio~\textsuperscript{1})            & I was able to localise and determine the position of where sounds were coming from.   \\  
\bottomrule
\end{tabularx}
}
\noindent{\footnotesize{\textsuperscript{1} These items were in the audio subscale in the original Presence Questionnaire; however this subscale was not utilised in our study.}}
\end{table}

Following the virtual experience we conducted an open-ended, semi-structured interview with the participants (outside of VR). This began with several questions to gather demographic data, before discussing the responses to the prototype, probing for further information as needed. A reflective approach was taken, showing the participants a screen recording of their experiences and how well they performed. Appendix~\ref{Appendix-questions} includes a list of indicative questions asked during the interview.

\subsubsection{Data Analysis Methods}

By triangulating both qualitative analysis of the interviews and quantitative analyses of the survey responses we aimed to gain a comprehensive understanding of user experiences within the virtual environments.

The MPS comprises three distinct subscales, each focusing on a specific aspect of presence: physical, social and self-presence. These subscales were used as the basis for our quantitative analysis, comparing differences between the experimental and control sound groups as well as between the spatial and non-spatial conditions. The two audio-specific questions were treated independently without being grouped into any subscale. 

Given that survey responses are inherently subjective and may not follow a normal distribution, it was necessary to apply nonparametric statistical tests for evaluating differences between samples. By employing the Mann--Whitney U test, we could assess potential differences in user experiences of presence and audio while accounting for variability in individual perceptions.

The qualitative data from the in-experience questions and the post-experience interview were transcribed and evaluated using thematic analysis~\parencite{Braun2006-js}. Using the `storybook' inductive approach allowed for a more flexible and constructionist examination of the data, which aligns well with design contexts where subjective interpretations are crucial to understanding user experiences. Through this approach, we identified patterns and themes emerging from participants' responses, seeking to reveal common trends, preferences, and concerns among users that would not be apparent through quantitative methods alone.

\section{Findings}

In this section, we present our quantitative results from the in-experience presence questionnaire, followed by our qualitative results from our thematic analysis of open-ended user responses.

As expected, users were not able to differentiate the intended behaviours of the AV in the control (no sound) group. This validated our experimental setup in which participants were asked to predict vehicle behaviour before seeing any indicative movement cues. While the efficacy of the specific vehicle sounds in helping users understand the intended vehicle behaviours is not the focus of this article, the lack of predictive power from the videos alone shows that the audio evaluation component of our methodology was (for lack of a better word) sound.

\subsection{Quantitative Results}\label{07-res-quant}

We conducted an assessment of normality in the quantitative data using the Shapiro--Wilk test due to the small sample sizes in our groups (control = 12; experiment = 36). This revealed that the data were not normally distributed, leading us to use the lower-powered but nonparametric Mann--Whitney U test to identify any differences between the samples.

\begin{figure}[H]

\centering
\includegraphics[width={\textwidth}]{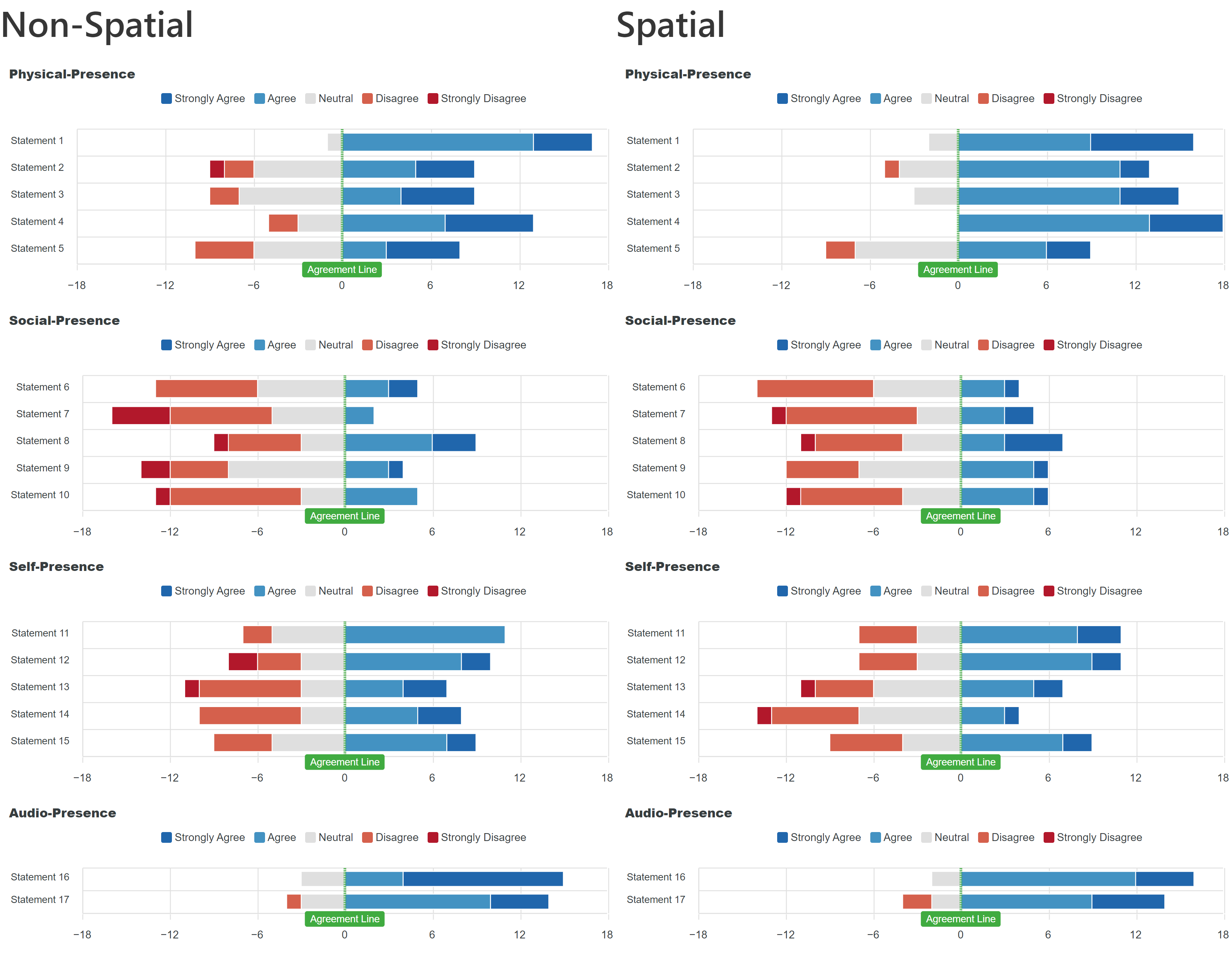}
\vspace{-10pt}
\caption{Visualisation of the responses for the Multimodal Presence Scale and additional audio questions, illustrating the levels of agreement among participants. These charts provide insights into the various dimensions of presence experienced by users during the study, highlighting particularly high physical presence.\label{07-fig4}}
\end{figure}

\begin{figure}[H]
\includegraphics[width={\textwidth}]{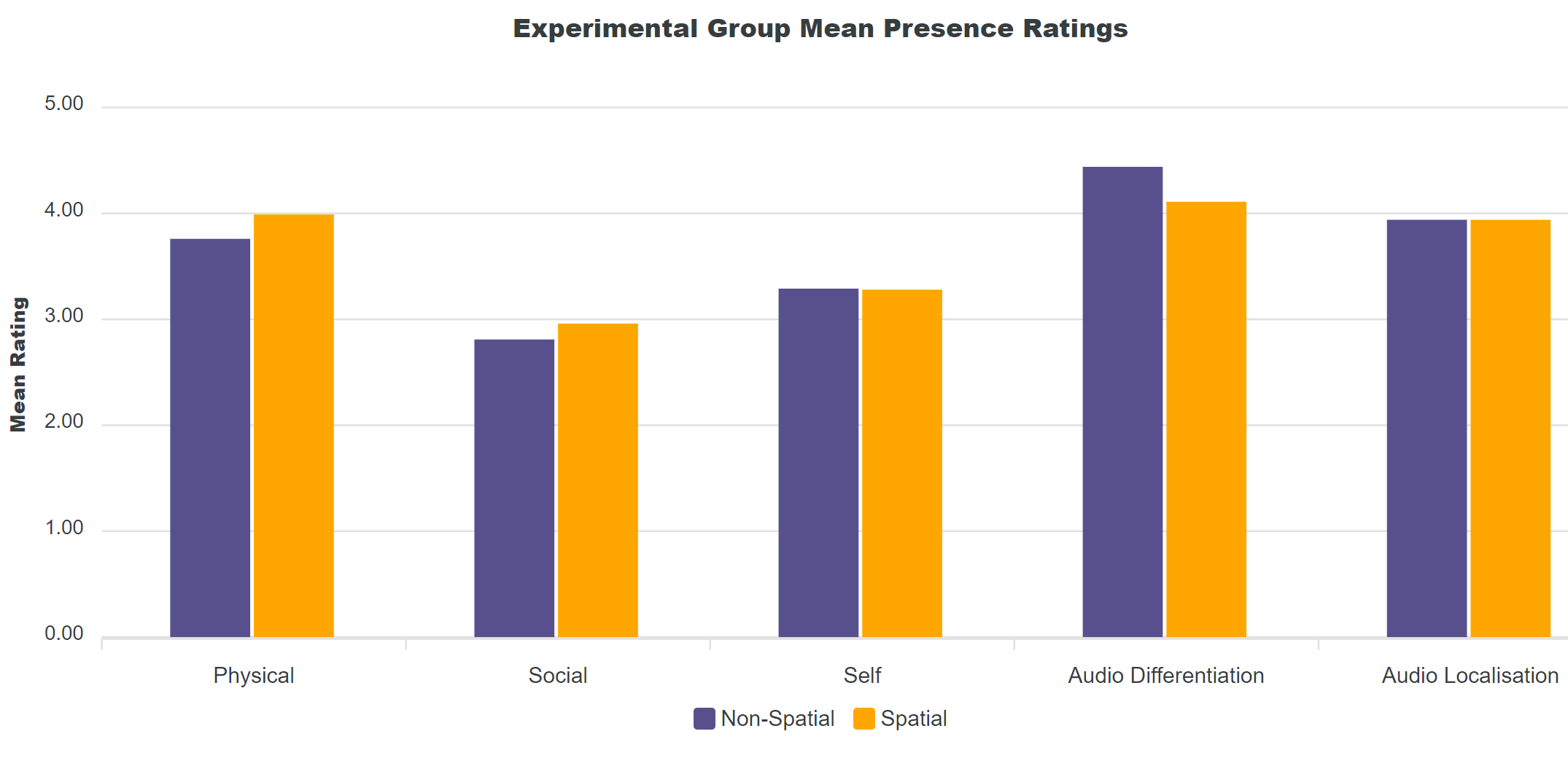}
\caption{Comparison of mean presence ratings from the Multimodal Presence Scale (MPS) subscales and additional audio questions for the experimental group. This shows no significant difference between the conditions, which indicates that the para-authentic stimuli in the virtual environment did not degrade users' feelings of presence, demonstrating their potential as ecologically valid platforms for Virtual Urban Field Studies.\label{07-fig5}}
\end{figure}

Within the experimental group, presence levels were found to be relatively high, particularly for the physical subscale (see Figure~\ref{07-fig4}). Specifically, for items 1 and 4 of the MPS, the total mean was greater than or equal to 4---indicating a high level of presence. This accords with the literature on 360-degree videos, which suggests that they can induce feelings of presence in VEs. No significant difference was observed between the spatial and non-spatial conditions for any of the MPS items or the additional audio questions (see Table~\ref{07-tab3} and Figure~\ref{07-fig5}). This indicates that the ambisonic spatialisation of ambient audio and vehicle sounds did not detectably influence presence levels. We were not able to detect an effect of age on these results (Mann--Whitney U test results, <= 25 vs. >26), with no significant difference between age groups in presence ratings (including audio-specific questions). However, due to the low median age, this test had low statistical power, making it infeasible to test the older group (>40) due to the small number of participants in this range.
\vspace{-10pt}
\vspace{-10pt}

\begin{table}[H] 
\centering
\caption{Differences between mean ratings from the Multimodal Presence Scale (MPS) subscales or the additional audio questions. No significant difference was observed between the non-spatial and spatial conditions in the experimental group with overlaid vehicle sounds, suggesting that the spatialisation of ambient audio and vehicle sounds did not impact presence.\label{07-tab3}}
   \setlength{\tabcolsep}{5.5mm}
\begin{tabularx}{15cm}{lllll}
\toprule
\textbf{Subscale/Item}	& \textbf{Non-Spatial}	& \textbf{Spatial} & \textbf{\emph{p} Value} & \textbf{Z Value}\\
\midrule
Physical Presence     & 3.76 & 3.99 & 0.3232 & $-0.9878$ \\
Social Presence       & 2.81 & 2.96 & 0.6794 & $-0.4132$ \\
Self Presence         & 3.29 & 3.28 & 0.9873 & 0.0159  \\
Audio Differentiation & 4.44 & 4.11 & 0.0937 & 1.6764  \\
Audio Localisation    & 3.94 & 3.94 & 0.8762 & $-0.1558$ \\
\bottomrule
\end{tabularx}
\end{table}

Comparing the control and the experimental groups, there was no significant difference between the presence ratings from the MPS and additional audio questions (see \mbox{Table~\ref{07-tab4}}). This suggests that the augmentation of the virtual environment with para-authentic virtual vehicle sounds did not detract from the users' feelings of presence, validating our sound design and production choices in integrating them. Subtle increases in scores for certain items may hint at the potential enhancement of presence due to the inclusion of vehicle sounds, although these observations were not statistically significant.

\begin{table}[H] 
\small
\caption{Comparison of control and experimental groups, showing no significant difference in mean presence ratings from the Multimodal Presence Scale (MPS) subscales and additional audio questions. This indicates that the para-authentic stimuli in the virtual environment did not degrade users' feelings of presence, demonstrating their potential as ecologically valid platforms for Virtual Urban Field Studies. \label{07-tab4}}
   \setlength{\tabcolsep}{2.6mm}
\begin{tabular}{lcccccc}
\toprule
& \multicolumn{3}{c}{\textbf{Control Group}} & \multicolumn{3}{c}{\textbf{Experimental Group}} \\
\midrule
\textbf{Survey Subscale/Item} & \textbf{Non-Spatial} & \textbf{Spatial} & \textbf{Total} & \textbf{Non-Spatial} & \textbf{Spatial} & \textbf{Total} \\\midrule
Physical Presence             & 3.43                 & 3.73             & 3.58              & 3.76                 & 3.99             & 3.86              \\
Social Presence               & 2.60                 & 2.87             & 2.73              & 2.81                 & 2.96             & 2.90              \\
Self Presence                 & 3.03                 & 2.70             & 2.87              & 3.29                 & 3.28             & 3.28              \\
Audio Differentiation         & 4.00                 & 3.50             & 3.75              & 4.44                 & 4.11             & 4.29              \\
Audio Localisation            & 3.50                 & 3.67             & 3.58              & 3.94                 & 3.94             & 3.94              \\
\bottomrule
\end{tabular}
\end{table}
\clearpage
\subsection{Qualitative Results}

The qualitative results from the thematic analysis are presented in Tables~\ref{07-tab5} and~\ref{07-tab6}, outlining the organising themes, corresponding themes, and codes derived from the participants' responses during the interview.

\begin{table}[H] 
\small
\caption{Organising themes and corresponding themes relating to virtual environments and feelings of embodiment in a Virtual Urban Field Study. Illustrative codes and occurrence counts, generated through thematic analysis.\label{07-tab5}}
        \centering\makebox[{\textwidth}][c]{%
        \renewcommand{\arraystretch}{1.5} 
        \begin{tabularx}{\textwidth}[t]{llXr}

        \toprule
        \textbf{Organising Theme}  & \textbf{Sub Theme}  & \textbf{Illustrative Codes}  & \textbf{Occurrences} \\
        \midrule
        \multirow[t]{12}{*}{Virtual Environments} & \multirow[t]{7}{*}{Realism}       & Felt like the real world                                    & 27 \\
                                               &                                & People were real and alive                                  & 16 \\
                                               &                                & People didn't feel real within the environment              & 15 \\
                                               &                                & Real world elements strengthened sense of presence   & 13 \\
                                               &                                & People walking around enhanced plausibility                   & 7  \\
                                               &                                & Feels like a video is being played                          & 6  \\
                                               &                                & Environment feels real whilst also feeling simulated        & 3  \\
                                               \cline{2-4}
                                               & \multirow[t]{3}{*}{Familiarity}& Familiarity with the location                               & 36 \\
                                               &                                & Real life experiences reinforced sense of being there & 5  \\
                                               &                                & People enhanced the familliar environment   & 2  \\
                                               \cline{2-4}
                                               & \multirow[t]{2}{*}{Interactivity} & Lack of social interaction with other people                & 15 \\
        \midrule
        \multirow[t]{18}{*}{Embodiment} & \multirow[t]{5}{*}{Motor Activation}    & Lack of interactivity impacted a sense of presence          & 13 \\
                                             &                                      & Not aware you could look around in the space                   & 4  \\
                                             &                                      & Lack of kinaesthetic feedback                                  & 3  \\
                                             &                                      & Unaware if they should use arms or look around   & 3  \\
                                             &                                      & Lack of locomotion                                             & 2  \\
                                             \cline{2-4}
                                             & \multirow[t]{3}{*}{Body Representation} & Not seeing hands causing lower sense of embodiment             & 12 \\
                                             &                                      & Arm wasn't projected into the scenarios                        & 5  \\
                                             &                                      & Lack of virtual representation of body                         & 4  \\
                                             \cline{2-4}
                                             & \multirow[t]{7}{*}{Hybrid Awareness}    & Virtual embodiment was an extension of their body              & 14 \\
                                             &                                      & Conscious of both internal and external worlds                 & 8  \\
                                             &                                      & External environment was still perceivable                     & 7  \\
                                             &                                      & Feels like they're operating something from the outside        & 5  \\
                                             &                                      & Felt that their real body was in the scene                     & 3  \\
                                             &                                      & Accuracy of height increased sense of embodiment               & 2  \\
                                             &                                      & Felt inorganically placed within the scene     & 2  \\
                                             \cline{2-4}
                                             & \multirow[t]{3}{*}{Reactions}           & Felt as if the AV was going to impact with them                & 18 \\
                                             &                                      & Feelings of self-preservation and concern of being hit         & 4  \\
                                             &                                      & Feeling real emotions made experience more real & 2 \\
        \bottomrule
        \end{tabularx}
    }
\end{table}

\begin{table}[H]
  \small
  \caption{Organising themes and corresponding themes relating to perception and response to audio in a Virtual Urban Field Study. Illustrative codes and occurrence counts, generated through thematic analysis.\label{07-tab6}}
  \centering\makebox[{\textwidth}][c]{%
  \renewcommand{\arraystretch}{1.5} 
    \begin{tabularx}{\textwidth}[t]{llXr}

      \toprule
      \textbf{Organising Theme}        & \textbf{Sub Theme}                 & \textbf{Illustrative Codes}                                & \textbf{Occurrences} \\
      \midrule
      \multirow[t]{11}{*}{Soundscapes} & \multirow[t]{5}{*}{Ambience}       & Generally could hear ambient sounds                        & 2                    \\
                                       &                                    & Background sounds of people were audible                   & 6                    \\
                                       &                                    & Background sounds made experience immersive                & 5                    \\
                                       &                                    & Background sounds felt natural                             & 3                    \\
                                       &                                    & Background sounds improved awareness of the environment    & 2                    \\
                                       \cline{2-4}
                                       & \multirow[t]{6}{*}{Spatialisation} & Ability to localise sounds improved immersion              & 9                    \\
                                       &                                    & Correctly recognised spatialising of audio                 & 6                    \\
                                       &                                    & Falsely perceived recognition of directional audio         & 4                    \\
                                       &                                    & Hard to recognise where sound were coming from             & 3                    \\
                                       &                                    & Spatial audio made distinguishing sound sources easier     & 2                    \\
                                       &                                    & Non-spatialised sound detracted from immersion             & 2                    \\
      \midrule
      \multirow[t]{8}{*}{Complexity}   & \multirow[t]{3}{*}{Focus}          & Lack of focus on background sounds                         & 22                   \\
                                       &                                    & Not hearing the consequential car sounds                   & 7                    \\
                                       &                                    & Background sound was a distraction to task                 & 3                    \\
                                       \cline{2-4}
                                       & \multirow[t]{5}{*}{Confusion}      & Vehicles with distinct sounds would be easily recognisable & 15                   \\
                                       &                                    & Confusion of AV sound with plane                           & 6                    \\
                                       &                                    & Sounds from multiple AVs can confuse pedestrians           & 2                    \\
                                       &                                    & Sound was too ambiguous                                    & 3                    \\
                                       &                                    & Context influences the interpretation of vehicle sounds    & 2                    \\
      \midrule
      \multirow[t]{8}{*}{Responses}    & \multirow[t]{3}{*}{Emotion}        & Rising sound escalated tension/intensity                   & 7                    \\
                                       &                                    & Loud/sudden sounds that are close can cause fear           & 5                    \\
                                       &                                    & Softer sounds are more calming and comfortable             & 4                    \\
                                       \cline{2-4}
                                       & \multirow[t]{5}{*}{Association}    & Existing sounds in vehicles                                & 33                   \\
                                       &                                    & Sounds very reminiscent of video game sounds               & 5                    \\
                                       &                                    & Sound {associated with/reminiscent of} particular environment & 3                \\
                                       &                                    & Sound was reminiscent of horror movies                     & 3                    \\
                                       &                                    & Reminiscent of emergency vehicle sounds                    & 3                    \\
      \bottomrule
    \end{tabularx}
  }
\end{table}

\subsubsection{360-Degree Video VR Environments}\label{07-res-VEs}

The majority of participants across all age-ranges found the 360-degree videos to be highly realistic and experienced them as (unsurprisingly) captures of real-life places and events. This sense of realism was further accentuated by the physical and material elements in the environment, solidifying a sense of scale through the presence of trees and architectural features. The visual phenomena, such as subtle movements in the trees and sunlight reflecting off glass buildings, also contributed to a sense of physical presence for some participants. The qualitative data suggest that this visual realism may explain the high agreement with statement 1 of the MPS across all groups and conditions (“The virtual environment seemed real to me”).

A significant number of users expressed their familiarity with the location (which was on our campus), an expected outcome given the sampling of participants from our University. This para-authenticity with a real-world environment seemed to facilitate the feeling of ‘being there’, in conjunction with the realism from the 360-degree videos. Even  the few participants who had not ever been on the university campus described how the environment was typical of an urban shared-use space, which provided them with a sense of familiarity. These feelings may potentially explain the relatively positive ratings for statement 4 of the MPS (“My experience in the virtual environment seemed consistent with my experiences in the real world”). 

Several users noted the lack of physical and social interaction within the environment, leading to a diminished sense of presence as the experience “felt like it was a video”. While acknowledging the environment’s realism, these users perceived the experience as a pre-recorded simulation (which---other than the vehicle sounds---it was). This made them feel like they were perceiving the situation as an outsider without any agency to control events. It is likely this partially contributed to the slightly lower ratings for statement 2 of the MPS (“I had a sense of acting in the virtual environment, rather than operating something from outside”). The ratings for statement 5 (“I was completely captivated by the virtual world”), might have also been lowered by the relatively short and repetitive nature of the task scenarios.

As anticipated (since there was no interaction with other humans), the social presence ratings were generally low, which was reinforced through the qualitative data. Participants had mixed opinions about the realism of other pedestrians in the scene, resulting in differing levels of agreement to statement 8 in the MPS (“The people in the virtual environment appeared to be sentient (conscious and alive) to me”). Participants who agreed with that statement reasoned that “having different people in different outfits just going about their day” felt real and that “they were minding their own business and everything feels natural”. Those who disagreed likened the other pedestrians to “non-player characters” or “bots”, who were not real because they had no awareness of the user and could not respond to their~actions.

\subsubsection{Embodiment: Perception, Awareness and Representation}\label{07-res-embodiment}

Several themes emerged concerning embodiment, which provided insights into users' personal experiences in the virtual environment that remained consistent irrespective of participants' age. One of the most prevalent themes was the lack of motor activation, referring to the limited amount of direct action with the body. This lack of motor activation likely contributed to the low self-presence ratings in the MPS. Factors such as the lack of locomotion and kinaesthetic feedback, as well as the limited need to look around (at least, once the car was spotted) may have contributed to this.

Another prevalent theme was the absence of a visual representation of the body or any of its parts. This likely impacted responses to statements 13 and 14 of the MPS, as users did not see their arms or hands during the virtual scenarios. Additionally, the limited body representation and motor activation may have influenced statement 2, resulting in users perceiving themselves as static observers rather than agents “acting in the virtual environment”.

Despite the limitations above, the qualitative data revealed some positive indications of embodiment and self-presence. A theme emerged regarding the hybrid awareness experienced by a few participants, reflecting a dual perception of the virtual and real worlds simultaneously. This sensation was expressed by participants as being “somewhere in the middle” or having “one foot in reality and one foot in virtual reality”. The relatively high agreement with statement 11 of the MPS (“I felt like my virtual embodiment was an extension of my real body within the virtual environment”) could be attributed to users acknowledging the involvement of their visual sensory apparatus in the virtual world despite the lack of bodily representation. This was summarised by one user who felt that “my body was here [in the real world] but I think my head and my eyes were more in the [virtual] environment”. 

For some users, this hybrid awareness was primarily characterised by a continued perception of the real world where they were “aware of their outside surroundings” while still engaging immersively with the virtual scenarios. However, for some this hybridity was not as positive for immersion, due to an ongoing awareness they were using VR and operating something from the outside. These kinds of users suggested “it still just did feel like I was doing stuff for the computer” and “I’m in the VR, this is not real, this is a virtual world”. It is possible this also contributed to the relatively low scores for statement 2 of the MPS (“I had a sense of acting in the virtual environment, rather than operating something from the outside”).

One of the most significant indications of embodiment was the participants' mental and, in some reported cases, physiological reaction to the apparent threat of the approaching vehicle. Even in the control group without vehicle sounds, the visual perception of the car driving directly towards the user elicited feelings of fear and nervousness. While the intensity of these sensations was probably less than what would be experienced in real life, and most participants did not “flinch” or “tense up”, most were able to “relate to actually being there” in that moment. This increased their sense of having a body in the virtual space rather than just watching a recording, further enhancing their feelings of presence.

\subsubsection{Ambient Soundscapes}\label{07-res-soundscapes}

Approximately half of the participants mentioned some aspect of the ambient audio in the soundscapes, including bird song, wind, pedestrians talking, planes and distant traffic sounds. Users felt that these sounds were “natural” and “accurate”, contributing to their feeling of being “immersed in the actual environment”. One user also claimed that this gave them a better sense of “spatial awareness”. In particular, the talking and plane sounds appeared to reinforce the familiarity with the environment (as discussed in Section~\ref{07-res-VEs}), likely strengthening the feelings of realism and physical presence. 

Participants’ recollections of the ambient audio were often vague or uncertain, with many referring only to “background noise” or identifying one or two distinct sounds within the overall soundscape. This was true for both the experimental conditions, suggesting the audio spatialisation did not make the ambient audio any more noticeable. When prompted, a few users did not perceive or remember any background audio at all (see Section~\ref{07-res-complexity}).

The qualitative data revealed mixed opinions regarding the spatialisation of the background audio. Some users in the spatial condition accurately perceived the sound as spatialised, specifying that “it all sounded like it was coming from an actual source that I could identify” (e.g., people talking to the left, or a plane flying overhead). In contrast, some of the users in the non-spatial condition falsely perceived localisation of the sounds, despite the audio being flattened to a single channel with no directionality. This occurred with the people talking (“I do remember a comment coming towards more of the right, I believe”) and the plane sound (“It sounded like it was coming from above me I guess”). These participants were generally less confident in the way they described perception than those in the spatial condition, even though no difference was visible in the quantitative data. This does not clarify why the non-spatial users perceived the directionality of the sounds, given the similarity in results for statement 16 of the presence questionnaire (see Section~\ref{07-res-quant}).

Some minor effects of the background audio were mentioned by individual participants. One person mentioned that the background audio distracted them from the task at hand. Another participant felt similarly, suggesting that too noisy an environment would make it harder to determine the vehicle position. No significant relationship was found between participants' age and the responses relating to the ambient soundscapes.

\subsubsection{User Responses to Para-Authentic Vehicle Sound Effects}\label{07-res-responses}

Several themes from our analysis characterised the participants’ responses to the designed sound effects for the AV. This is a positive indication of the efficacy of para-authentic virtual sounds embedded within 360$^\circ$ video as an approach, although analysing the comparative efficacy of specific sound designs is out of scope for this article.

One thing that emerged from our analysis was the highly varied emotional response to sounds, reinforcing the subjective nature of individual reactions to auditory stimuli. Negative emotions were associated with certain sounds, such as the fear caused by sudden sounds or the “escalation of tension” resulting from rising pitch. An increase in volume and deeper bass were also linked to the feeling that “something bad was going to happen”. Conversely, some participants experienced positive emotional responses to specific sounds. Softer sounds were perceived as more “calm” and “comfortable”, while more melodic sounds were described as “passive and non-confrontational”. Additionally, one user referred to the high speed of some sounds as “exciting”. All of these responses suggest that sounds can prompt different “emotional responses based on [their] intensity”.

Participants also drew associations with sounds they had encountered in the past. The most common comparisons were with existing vehicle sounds, such as engine noises, horns and sirens in emergency vehicles, as well as warning beeps for systems such as reversing sensors or pedestrian detection. These familiar sounds seemed to provide appropriate mental models for participants, increasing familiarity and potentially supporting presence. Older participants (>26) exclusively referenced existing vehicle sounds, while younger participants (<26) also made reference to media, such as sounds from films and video games. Some of these participants suggested that the higher pitch sounds reminded them of the “stabbing”-style soundtracks used in classic horror movies, contributing to the fear response mentioned previously. Lastly, a few participants associated the sounds with non-vehicular warning systems, such as fire alarms. 

\subsubsection{Sound Complexity and Ambiguity}\label{07-res-complexity}
Another organising theme highlighted the intricate nature of sounds and the challenges that a dynamic real-world setting can pose to communicating clear meaning. Some participants experienced confusion regarding the source and characteristics of certain sounds. For example, some people were unable to distinguish between the para-authentic virtual vehicle sounds and the noise of a plane passing overhead. In a more traditional HCI user study context, it might be prudent to remove incidental interruptive noises such as aeroplanes from experiments; the challenge of VUFS prototyping is that such infrequent-but-loud noises are part of the urban soundscape. While the plane and the car sounds did conflict, the possibility of them doing so both shows the realism and fidelity of our overlaid sounds and suggests a challenge for AV sound designers to address. Despite potential differences arising from age (e.g., hearing ability), responses about confusion resulting from ambiguity did not appear to be influenced by these variations.

\section{Discussion}

Here we discuss the two key contributions of this study: the use of presence while developing VUFS, and the use of VUFS more broadly as a way to evaluate urban interaction designs.

\subsection{Presence as a Proxy for Verisimilitude in VUFS}\label{07-dis-presence}

The first objective of this research was to explore the  efficacy and characteristics of VUFS (particularly with a sound design focus) as a way to approximate  field studies. We found that the notion of presence serves as a valuable indicator of verisimilitude---by which we mean the capacity of the virtual environment to stand in for a real field study---for VUFS. When conducting virtual field studies, it is essential not only to evaluate the effectiveness of a prototype in achieving its intended goals (design efficacy), but also to measure how well the prototype actually reflects the intended environment (VUFS efficacy/verisimilitude). In some HCI studies that utilise virtual environments, presence is clearly desirable but neither measured nor explicitly designed for. By making presence an explicit goal of our iterative design process, we aimed to maximise the presence dimensions that were relevant for our specific context-based interface prototype.

The significance of maximising the various dimensions of presence depends on the specific application and context in which it is employed. For instance, architectural contexts might necessitate a high level of physical presence, ensuring that users have an authentic perception of factors such as space and scale. By contrast, social robotics applications would likely benefit from enhanced social presence, fostering a more realistic relationship between the user and the robot. Wearable technology might rely on high self-presence, where the user experiences a strong sense of embodiment and connection with the device. In urban settings, particularly when designing technologies for automated cities, there may be good reasons why all three dimensions of presence become relevant.

The sense of physical presence in our study was primarily achieved through realistic recordings and familiarity with the environment contributing to both a ``place illusion'' and ``plausibility illusion''~\parencite{Slater2009-iz}. The quantitative results indicate that the fidelity of the virtual environment was sufficient to create a sense of physical presence for many users. Although some users mentioned the blurriness of the video due to resolution limitations, it did not seem to strongly impact the realism or sense of ‘being there’. Participants not only felt they were in a real place, but also believed that the scenarios could genuinely have occurred. This plausibility was predominantly evidenced in the qualitative data, where users perceived realism through audiovisual recording fidelity and comparisons to their past experiences.

Participants who had frequently visited the campus exhibited a specific familiarity with the space used in recordings, while the few that had never visited still reported a general familiarity with similar spaces (i.e., shared pedestrian/vehicle urban areas). Due to the limitations of our recruitment methods, most of our participants were well-acquainted with this environment. Despite this high level of familiarity, our users still found the experience realistic and plausible, which is a good indication of the validity of the method (as this group would be more likely to find potentially presence-breaking inconsistencies in the virtual experience). Given this limitation, our findings are most applicable to the context of on-campus AVs (such as for mobility), rather than more general AV use-cases. Designers of future VUFS should consider the levels of both specific and general familiarity which are appropriate for their use cases. For example, in a study on augmented-reality-facilitated navigation in urban environments, general (but not specific) familiarity may be desirable, to prevent effects on task efficacy.

The lack of interaction with the environment may have slightly detracted from physical presence. However, this did not appear to negatively impact the study, as users were still primarily focused on the vehicle and were able to engage with the scenarios presented. Design contexts which involve more movement or physical action must carefully consider these interactions to successfully leverage context-based interface prototypes. In these cases, it may be better to leverage alternative approaches to creating VEs (such as 3D computer generated worlds), for more complex interactivity.

Despite the lack of social interactions, other pedestrians in the scenarios contributed to the plausibility, likely somewhat enhancing social presence. Limited embodiment resulting from minimal physical engagement and the absence of a visual representation of the body may have impacted self-presence in relation to motor function. Nevertheless, there was still some level of self-presence at a sensory and cognitive level, further validating the use of this prototyping method for conducting studies of user perceptions.

\subsection{Virtual Urban Field Studies for Evaluating Interaction Design}\label{07-dis-vufs}

Our study supports the notion that VR is a highly suitable medium for simulating field studies testing interaction design, supporting what has been suggested before in similar studies~\parencite{Makela2020-mx}. This is particularly relevant for situations that could be considered dangerous, such as our case of an AV driving towards a pedestrian with the possibility of collision. 

The quantitative data showed there was no significant deterioration in feelings of presence when para-authentic virtual sounds were used to augment the real ambient soundscape recordings. There was actually a small (although not significant) increase in all subscale ratings as well as most individual items of the MPS in the experimental group when compared to the control, potentially indicating increased presence due to the layered sounds. This could partly be due to emotional responses to the sound (discussed below and further in Section~\ref{07-dis-authentic}) making people feel more present. Perhaps users suspended their disbelief upon hearing sounds from the car, further immersing themselves in the scenarios without critically examining the unreal elements. We believe this is an area that warrants further investigation in context-based interface prototypes, including other sensory contexts beyond sound---what is the impact on presence of elements in the environment seeming to “respond to” or “address” the participant?

When the para-authentic virtual vehicle sounds were overlaid on the recorded ambient soundscape there was no significant difference in presence between the spatial and non-spatial conditions. This is actually a positive, as it suggests that for these kinds of experiments, a simple level of spatialisation is enough to generate presence and appropriate task outcomes. Ambisonic recording hardware and software is expensive, and requires specialised skillsets to operate. For VUFS developers not to have to engage with it is a benefit, leaving them free to focus on the design of the sounds themselves, rather than on sound production. 

As suggested by \textcite{Jain2021-ab}, “designers make creative decisions for sounds in VR to be more pleasant and manageable”. In our design process, we iterated the vehicle sounds multiple times to test different characteristics and how they interacted with the virtual environment. This resulted in tweaking qualities such as pitch, tempo, timbre and melody. We tuned features such as positionality and reverb for simplicity not realism (mostly to avoid having to take detailed acoustic scans of the environment). For example, the attenuation of sound in the space was not physically realistic, but was apparently good enough to situate the sound well enough that users perceived no difference in their sense of “being there”, even in the audio-specific questions.

Background ambient soundscapes played a crucial role in the overall cohesiveness of the virtual environment. As one participant stated, “it's easier to report about an experience that feels off than about an experience that is natural”, suggesting they would have noticed if the sounds were not appropriate. This aligns with prior research~\parencite{Kern2020-cm}, demonstrating that including a plausible soundscape has a high impact on presence and perceptions of realism. For the ambient audio, the level of spatialisation appeared not to significantly impact feelings of presence. In fact, the layering of para-authentic virtual sounds caused users to focus less on or even completely ignore the background audio. It is possible a lower-order (thus cheaper and more convenient to use) microphone could have been utilised, without sacrificing the presence-inducing effects of having an appropriate soundscape. 

\subsection{Generating Authentic User Responses with Para-Authentic Stimuli}\label{07-dis-authentic}

The second objective of this research was to investigate the impact of overlaying virtual sounds on recorded imagery on user experience in VUFS. Our qualitative results suggest that the characteristics of the para-authentic vehicle sounds did influence users’ responses to the virtual stimuli. For example, some participants claimed to feel fear and nervousness in response to the intense crescendos with rising pitch and oscillation. This anticipatory fear reaction is consistent with the literature, which states that “individuals make behavioural , cognitive, and physiological adjustments that facilitate the process of coping with the upcoming stressor”~\parencite{Pulopulos2020-ls}. The users did not get to see the impact of the vehicle before making their decisions as the video was paused before the outcome was shown, yet these sensations still arose. That VR could generate these anticipatory stress responses, even in the absence of any real danger, supports the efficacy of VUFS for these kinds of experiments. Other users reported feeling a sense of calmness from the more melodic audio patterns with steady rhythms. Both of these responses indicate the ecological validity of our VEs, where the users are affectively engaging with the sensory stimuli as if they were real. Our results reinforce that sounds can influence cognitive processing and potential emotional responses by leveraging associations. These associations can be both domain-specific (e.g., similar sounds in vehicles) or draw on outside references (e.g., game sound effects). Our qualitative data reinforce that our users interpreted our sound designs to impart meaning and (potentially) emotional responses, often drawing on past experiences (e.g., horn-like sounds captured attention and then induced caution/alertness) to do so.

When designing a VUFS prototype to maximise physical presence, sounds intended to generate a response should adhere to certain guidelines regarding realism. They should at least be somewhat spatialised to situate them within the environment, with some directionality to enhance plausibility. In some cases, higher-order ambisonics may offer additional advantages that offset the increased effort, specialised equipment, and expert knowledge required. The sounds should support existing mental models, as overly fantastical and inappropriate sounds may detract from the plausibility. Our study focused on notification-style communication in small electric AVs, as opposed to simulated motor sounds \parencite[c.f.][]{Moore2020-be}. Both kinds of sounds require further research to explore potential sound signatures for different vehicles (considering factors such as vehicle size, velocity and pedestrian proximity).

Ensuring a plausible and appropriate multisensory environmental context is crucial for inducing authentic user behaviours within virtual settings. For instance, participants may have perceived our car sounds very differently if they did not originate from an approaching vehicle, and instead were presented without video or background audio. Incorporating realistic auditory stimuli within virtual environments effectively elicits accurate user responses and enhances the overall immersive experience.

\subsection{Considerations for Designing Immersion-Supporting Sounds}\label{07-dis-sounds}
While ambient soundscapes are often designed to be unnoticeable (or at least unobtrusive), there are instances where these sounds may attract the user’s attention and further ground their experience. While the background sounds in our study were recorded, it could benefit some VUFS to utilise composed soundscapes. As we discovered in our study, care should be taken to not distract the user from the task at hand, perhaps by synchronising the immersion-supporting background sounds with specific lulls in the scenarios. This could help to sustain immersion in both the physical and fictional elements of a virtual environment, potentially preventing users from becoming distracted when not actively engaged in a task. These sounds should align with the visuals and be appropriately calibrated for the scene (i.e., if a vehicle noise is heard, the vehicle should be visible somewhere).

When designing interface sounds, designers should also consider possible confusion of sound sources and characteristics. In our study, one of the scenarios contained the audio of a plane flying overhead, which coincided with the layered para-authentic vehicle sound. Although we considered the potential of overlapping sound frequencies in our design process, we did not account for the general similarity of our sound with other environmental sounds in the environment (such as a gradual decrease in pitch indicating the AV slowing down). This led to an interesting situation in which participants confused and sometimes conflated the vehicle sound with the plane sound. This reinforces the notion of utilising sounds with prior associations to leverage users’ potential existing mental models (see Section~\ref{07-res-responses}). However, it is essential to design these sounds in a way that minimises any negative interaction between audio interfaces and background sounds. Striking a balance between realism and experiment design is essential in designing effective auditory stimuli for VUFS in particular. Retaining realistic elements, such as the sound of a plane, contributes to the immersive experience, but may also increase the scope of design issues that might be identifiable. For example, if designing sounds for vision-impaired users at bus stops, it would be valuable to know if those sounds might be confused with a bus engine. The same would be true if LED lights on a vehicle interface were confused with reflections from other light sources. Ultimately, our opinion is that VUFS verisimilitude should win out, as the rigour of controlled experiments conducted as part of the evaluation of a design prototype should be secondary to the design quality of that prototype. That said, one should be just as careful making claims about a VUFS experiment that lacked certain environmental controls as a real-world experiment with the same issue.

\section{Conclusions}
In this study, we explored the potential of virtual reality for testing interactive systems, specifically focusing on the context of AV--pedestrian interaction in shared spaces. We proposed Virtual Urban Field Studies (VUFS) as the label for studies using context-based interface prototypes to approximate the context of urban interaction. Engaging in research-through-design, we developed a prototype for AV auditory interfaces, which we evaluated with a user study. Our prototype employed 360-degree video and spatial audio recordings to create a virtual environment (VE), then augmented it by overlaying para-authentic vehicle interface sounds. 

We also examined the role of presence in determining the verisimilitude and real-world applicability of the simulated urban environments. By investigating how the dimensions of presence shaped user experiences in VUFS, we identified ways that the immersive affordances (both perceptual and psychological) of VR can be leveraged to provide an ecologically valid testing environment. This was quantitatively validated by our data showing that the addition of para-authentic vehicle sounds led to no deterioration in~presence.

The results of our study indicated a high level of physical presence in our virtual experience, partly due to the realism of recorded 360-degree video and spatial audio soundscapes. This was represented in the quantitative data (see Section~\ref{07-res-quant}) as well as the qualitative data (see Section~\ref{07-res-VEs}). This was true whether the participant experienced the “spatial” audio or the simple stereo version, suggesting that expensive and specialised ambisonic technology is not necessary for VUFS development, at least not when sound localisation is not the focus. We did not find any effects of age on the efficacy of VUFS, but we acknowledge the limited power of our study due to a lack of older users (only 8\% were over 40). Additionally, our recruitment from a university setting might have introduced a bias towards  higher levels of education and socioeconomic status, thus more exposure to new technologies such as VR. Further studies would be required to assess the viability and effectiveness of the VUFS approach across a more representative sample of urban~inhabitants.

Physical presence was also bolstered by the experience of a “place illusion” (resulting from familiar and authentic locations) and “plausibility illusion” (resulting from scenarios that could genuinely occur). Participants reported feelings of social and self-presence, just not as high as physical presence. While there was no direct interaction with other social actors in the experience, having other pedestrians in the recordings contributed to the plausibility of the scenarios. The moderate ratings for self-presence suggest that minimal embodiment can be experienced from just experiencing 360-degree sensory input, even without bodily representation in the VE. 

Future research should explore the use of VUFS to explore different applications for urban technology. Use-cases such as pedestrian interaction with AVs are particularly well suited, given the risk and costs associated with the vehicles. However, we believe this method can be more generally applied for the design and evaluation of many other urban interactions and interfaces.

We have discussed the advantages and potential disadvantages of employing VEs, using approaches such as 360-degree video and spatial audio. Based on our findings, we have offered design considerations for operationalising presence for the purpose of VUFS. We hope this method can reduce the time, cost and development effort associated with traditional field studies. By harnessing the power of VEs and measuring presence as a proxy for verisimilitude, designers and researchers can create more engaging and authentic experiences, advance urban HCI, and develop innovative solutions for real-world urban~challenges.

\appendix
\section{Appendix}
\label{Appendix-questions}
\begin{itemize}
    \item What do you remember focusing on during this part of the experience?
    \item Comparing that scenario to the last scenario, how do you think they were similar or different?
    \item Why did you agree/disagree with that statement from the questionnaire?
    \item What were you hearing during the experience?
    \item What kind of sounds were you able to distinguish?
    \item Were there any sounds that stand out in your mind? Either on the positive or the negative side?
    \item Was there anything about the sound that either contributed to immersion or your relationship with the vehicle?
    \item What kind of cues were you getting that helped you localise the sound?
    \item Was there anything that felt consistent/inconsistent with your real-world past experiences?
    \item Can we dive further into what about this experience made it feel real to you?
    \item What were you thinking about the other people around in the environment?
    \item How do you think that would compare if it was a more stylised rendered 3D representation with virtual characters?
    \item How did you feel about just generally physically being inside that experience?
    \item Which parts of your body felt involved inside that experience?
    \item What kind of things were you thinking were happening to your body?
\end{itemize}

\printbibliography

@INPROCEEDINGS{Zimmerman2007-cc,
  title     = "Research Through Design As a Method for Interaction Design
               Research in {HCI}",
  author    = "Zimmerman, John and Forlizzi, Jodi and Evenson, Shelley",
  booktitle = "Proceedings of the SIGCHI Conference on Human Factors in
               Computing Systems",
  publisher = "ACM",
  address   = "New York, NY, USA",
  pages     = "493--502",
  series    = "CHI '07",
  year      =  2007
}

@ARTICLE{Slater2009-iz,
  title    = "Place illusion and plausibility can lead to realistic behaviour in
              immersive virtual environments",
  author   = "Slater, Mel",
  journal  = "Philos. Trans. R. Soc. Lond. B Biol. Sci.",
  volume   =  364,
  number   =  1535,
  pages    = "3549--3557",
  month    =  dec,
  year     =  2009,
  language = "en"
}

@ARTICLE{Lee2004-qg,
  title    = "Presence, Explicated",
  author   = "Lee, Kwan Min",
  journal  = "Commun. Theory",
  volume   =  14,
  number   =  1,
  pages    = "27--50",
  month    =  feb,
  year     =  2004
}

@ARTICLE{Braun2006-js,
  title   = "Using thematic analysis in psychology",
  author  = "Braun, Virginia and Clarke, Victoria",
  journal = "Qual. Res. Psychol.",
  volume  =  3,
  number  =  2,
  pages   = "77--101",
  month   =  jan,
  year    =  2006
}

@ARTICLE{Witmer1998-fz,
  title     = "Measuring Presence in Virtual Environments: A Presence
               Questionnaire",
  author    = "Witmer, Bob G and Singer, Michael J",
  journal   = "Presence: Teleoperators and Virtual Environments",
  publisher = "MIT Press",
  volume    =  7,
  number    =  3,
  pages     = "225--240",
  month     =  jun,
  year      =  1998
}

@ARTICLE{Gilbert2016-wf,
  title     = "Perceived Realism of Virtual Environments Depends on Authenticity",
  author    = "Gilbert, Stephen B",
  journal   = "Presence: Teleoperators and Virtual Environments",
  publisher = "MIT Press",
  volume    =  25,
  number    =  4,
  pages     = "322--324",
  month     =  dec,
  year      =  2016
}

@ARTICLE{Biocca1997-fx,
  title     = "The Cyborg's Dilemma: Progressive Embodiment in Virtual
               Environments",
  author    = "Biocca, Frank",
  journal   = "J. Comput. Mediat. Commun.",
  publisher = "Oxford University Press",
  volume    =  3,
  number    =  2,
  month     =  sep,
  year      =  1997
}

@ARTICLE{Keyson2009-mx,
  title   = "Empirical Research Through Design",
  author  = "Keyson, David V and Alonso, Miguel Bruns",
  journal = "Proceedings of the 3rd IASDR Conference on Design Research",
  pages   = "4548--4557",
  month   =  oct,
  year    =  2009
}

@ARTICLE{Makransky2017-kw,
  title     = "Development and Validation of the Multimodal Presence Scale for
               Virtual Reality Environments: A Confirmatory Factor Analysis and
               Item Response Theory Approach",
  author    = "Makransky, Guido and Lilleholt, Lau and Aaby, Anders",
  journal   = "Comput. Human Behav.",
  publisher = "Elsevier",
  volume    =  72,
  month     =  feb,
  year      =  2017
}

@ARTICLE{Grassini2020-wu,
  title    = "Questionnaire Measures and Physiological Correlates of Presence: A
              Systematic Review",
  author   = "Grassini, Simone and Laumann, Karin",
  journal  = "Front. Psychol.",
  volume   =  11,
  pages    =  349,
  month    =  mar,
  year     =  2020,
  language = "en"
}

@ARTICLE{Felton2022-np,
  title     = "Presence: A Review",
  author    = "Felton, William M and Jackson, Russell E",
  journal   = "International Journal of Human–Computer Interaction",
  publisher = "Taylor \& Francis",
  volume    =  38,
  number    =  1,
  pages     = "1--18",
  month     =  jan,
  year      =  2022
}

@INCOLLECTION{Putze2020-il,
  title     = "Breaking The Experience: Effects of Questionnaires in {VR} User
               Studies",
  author    = "Putze, Susanne and Alexandrovsky, Dmitry and Putze, Felix and
               Höffner, Sebastian and Smeddinck, Jan David and Malaka, Rainer",
  booktitle = "Proceedings of the 2020 CHI Conference on Human Factors in
               Computing Systems",
  publisher = "Association for Computing Machinery",
  address   = "New York, NY, USA",
  pages     = "1--15",
  month     =  apr,
  year      =  2020
}

@BOOK{Hale_KS_Stanney_KM2014-rp,
  title     = "Handbook of Virtual Environments: Design, Implementation, and
               Applications",
  editor    = "{Hale, K.S., \& Stanney, K.M.}",
  publisher = "CRC Press",
  edition   = "Second Edition",
  year      =  2014
}

@ARTICLE{Souza2021-hk,
  title     = "Measuring Presence in Virtual Environments: A Survey",
  author    = "Souza, Vinicius and Maciel, Anderson and Nedel, Luciana and
               Kopper, Regis",
  journal   = "ACM Comput. Surv.",
  publisher = "Association for Computing Machinery",
  address   = "New York, NY, USA",
  volume    =  54,
  number    =  8,
  pages     = "1--37",
  month     =  oct,
  year      =  2021
}

@ARTICLE{Nilsson2016-xl,
  title     = "Immersion Revisited: A review of existing definitions of
               immersion and their relation to different theories of presence",
  author    = "Nilsson, Niels Christian and Nordahl, Rolf and Serafin, Stefania",
  journal   = "Hum. Technol. Interdiscip. J. Hum. ICT Environ.",
  publisher = "Centre of Sociological Research, NGO",
  volume    =  12,
  number    =  2,
  pages     = "108--134",
  month     =  nov,
  year      =  2016
}

@ARTICLE{Riva2004-wm,
  title    = "The layers of presence: a bio-cultural approach to understanding
              presence in natural and mediated environments",
  author   = "Riva, Giuseppe and Waterworth, John A and Waterworth, Eva L",
  journal  = "Cyberpsychol. Behav.",
  volume   =  7,
  number   =  4,
  pages    = "402--416",
  month    =  aug,
  year     =  2004,
  language = "en"
}

@INPROCEEDINGS{Makela2020-mx,
  title     = "Virtual Field Studies: Conducting Studies on Public Displays in
               Virtual Reality",
  author    = "Mäkelä, Ville and Radiah, Rivu and Alsherif, Saleh and Khamis,
               Mohamed and Xiao, Chong and Borchert, Lisa and Schmidt, Albrecht
               and Alt, Florian",
  booktitle = "Proceedings of the 2020 CHI Conference on Human Factors in
               Computing Systems",
  publisher = "Association for Computing Machinery",
  address   = "New York, NY, USA",
  pages     = "1--15",
  series    = "CHI '20",
  month     =  apr,
  year      =  2020
}

@INPROCEEDINGS{Moore2020-be,
  title     = "Sound Decisions: How Synthetic Motor Sounds Improve Autonomous
               Vehicle-Pedestrian Interactions",
  author    = "Moore, Dylan and Currano, Rebecca and Sirkin, David",
  booktitle = "12th International Conference on Automotive User Interfaces and
               Interactive Vehicular Applications",
  publisher = "Association for Computing Machinery",
  address   = "New York, NY, USA",
  pages     = "94--103",
  series    = "AutomotiveUI '20",
  month     =  sep,
  year      =  2020
}

@INPROCEEDINGS{Nguyen2019-sn,
  title     = "Designing for projection-based communication between autonomous
               vehicles and pedestrians",
  author    = "Nguyen, Trung Thanh and Holländer, Kai and Hoggenmueller, Marius
               and Parker, Callum and Tomitsch, Martin",
  booktitle = "Proceedings of the 11th International Conference on Automotive
               User Interfaces and Interactive Vehicular Applications",
  publisher = "ACM",
  address   = "New York, NY, USA",
  month     =  sep,
  year      =  2019,
  language  = "en"
}

@INPROCEEDINGS{Johansen2022-zg,
  title     = "Characterising Soundscape Research in Human-Computer Interaction",
  author    = "Johansen, Stine S and van Berkel, Niels and Fritsch, Jonas",
  booktitle = "Designing Interactive Systems Conference",
  publisher = "Association for Computing Machinery",
  address   = "New York, NY, USA",
  pages     = "1394--1417",
  series    = "DIS '22",
  month     =  jun,
  year      =  2022
}

@ARTICLE{Bertet2006-kn,
  title     = "{3D} sound field recording with higher order ambisonics.
               Objective measurements and validation of a {4th} order spherical
               microphone",
  author    = "Bertet, Stéphanie and Daniel, Jérôme and Moreau, Sébastien",
  journal   = "Proceedings of 120th AES Convention",
  publisher = "pcfarina.eng.unipr.it",
  month     =  jan,
  year      =  2006,
  language  = "en"
}

@ARTICLE{Fellgett1974-sl,
  title    = "Ambisonic reproduction of directionality in surround-sound systems",
  author   = "Fellgett, P B",
  journal  = "Nature",
  volume   =  252,
  number   =  5484,
  pages    = "534--538",
  month    =  dec,
  year     =  1974
}

@ARTICLE{Hong2017-gw,
  title     = "Spatial Audio for Soundscape Design: Recording and Reproduction",
  author    = "Hong, Joo Young and He, Jianjun and Lam, Bhan and Gupta, Rishabh
               and Gan, Woon-Seng",
  journal   = "NATO Adv. Sci. Inst. Ser. E Appl. Sci.",
  publisher = "Multidisciplinary Digital Publishing Institute",
  volume    =  7,
  number    =  6,
  pages     =  627,
  month     =  jun,
  year      =  2017,
  language  = "en"
}

@ARTICLE{Waller2021-pd,
  title    = "Meditating in Virtual Reality 3: {360°} Video of Perceptual
              Presence of Instructor",
  author   = "Waller, Madison and Mistry, Divya and Jetly, Rakesh and Frewen,
              Paul",
  journal  = "Mindfulness",
  volume   =  12,
  number   =  6,
  pages    = "1424--1437",
  month    =  mar,
  year     =  2021,
  language = "en"
}

@ARTICLE{Steuer1992-at,
  title     = "Defining Virtual Reality: Dimensions Determining Telepresence",
  author    = "Steuer, Jonathan",
  journal   = "J. Commun.",
  publisher = "John Wiley \& Sons, Ltd",
  volume    =  42,
  number    =  4,
  pages     = "73--93",
  month     =  dec,
  year      =  1992
}

@ARTICLE{Ijsselsteijn2000-rf,
  title     = "Presence: Concept, determinants and measurement",
  author    = "Ijsselsteijn, Wijnand A and de Ridder, Huib and Freeman, Jonathan
               and Avons, Steve E",
  journal   = "Proceedings of SPIE - The International Society for Optical
               Engineering",
  publisher = "Society of Photo-optical Instrumentation Engineers",
  volume    =  3959,
  month     =  nov,
  year      =  2000
}

@INCOLLECTION{McMahan2013-uv,
  title     = "Immersion, engagement, and presence: A method for analyzing 3-{D}
               video games",
  author    = "McMahan, Alison",
  booktitle = "The video game theory reader",
  publisher = "Routledge",
  pages     = "67--86",
  year      =  2013
}

@ARTICLE{Kern2020-cm,
  title    = "Audio in {VR}: Effects of a Soundscape and Movement-Triggered Step
              Sounds on Presence",
  author   = "Kern, Angelika C and Ellermeier, Wolfgang",
  journal  = "Front Robot AI",
  volume   =  7,
  pages    =  20,
  month    =  feb,
  year     =  2020,
  language = "en"
}

@INPROCEEDINGS{Jain2021-ab,
  title     = "A Taxonomy of Sounds in Virtual Reality",
  author    = "Jain, Dhruv and Junuzovic, Sasa and Ofek, Eyal and Sinclair, Mike
               and Porter, John and Yoon, Chris and Machanavajhala, Swetha and
               Ringel Morris, Meredith",
  booktitle = "Designing Interactive Systems Conference 2021",
  publisher = "Association for Computing Machinery",
  address   = "New York, NY, USA",
  pages     = "160--170",
  series    = "DIS '21",
  month     =  jun,
  year      =  2021
}

@INPROCEEDINGS{Flohr2020-vq,
  title     = "Context-Based Interface Prototyping and Evaluation for (Shared)
               Autonomous Vehicles Using a Lightweight Immersive Video-Based
               Simulator",
  author    = "Flohr, Lukas A and Janetzko, Dominik and Wallach, Dieter P and
               Scholz, Sebastian C and Krüger, Antonio",
  booktitle = "Proceedings of the 2020 ACM Designing Interactive Systems
               Conference",
  publisher = "Association for Computing Machinery",
  address   = "New York, NY, USA",
  pages     = "1379--1390",
  series    = "DIS '20",
  month     =  jul,
  year      =  2020
}

@INPROCEEDINGS{Hoggenmuller2021-ae,
  title     = "Context-Based Interface Prototyping: Understanding the Effect of
               Prototype Representation on User Feedback",
  author    = "Hoggenmüller, Marius and Tomitsch, Martin and Hespanhol, Luke and
               Tran, Tram Thi Minh and Worrall, Stewart and Nebot, Eduardo",
  booktitle = "Proceedings of the 2021 CHI Conference on Human Factors in
               Computing Systems",
  publisher = "Association for Computing Machinery",
  address   = "New York, NY, USA",
  number    = "Article 370",
  pages     = "1--14",
  series    = "CHI '21",
  month     =  may,
  year      =  2021
}

@ARTICLE{Lombard1997-lq,
  title     = "At the Heart of It All: The Concept of Presence",
  author    = "Lombard, Matthew and Ditton, Theresa",
  journal   = "J. Comput. Mediat. Commun.",
  publisher = "John Wiley \& Sons, Ltd",
  volume    =  3,
  number    =  2,
  month     =  sep,
  year      =  1997
}

@INPROCEEDINGS{Murphy2014-vk,
  title     = "Spatial impulse response measurement in an urban environment",
  author    = "Murphy, Damian Thomas and Stevens, Francis Kit Murfin",
  booktitle = "Spatial Audio, AES 55th International Conference",
  address   = "FIN",
  pages     =  8,
  month     =  aug,
  year      =  2014,
  language  = "en"
}

@INPROCEEDINGS{Poeschl-Guenther2013-em,
  title     = "Integration of spatial sound in immersive virtual environments an
               experimental study on effects of spatial",
  author    = "Poeschl-Guenther, Sandra and Wall, Konstantin and Döring, Nicola",
  booktitle = "Virtual Reality (VR), 2013 IEEE",
  publisher = "unknown",
  pages     = "129--130",
  month     =  mar,
  year      =  2013
}

@ARTICLE{Potter2022-fa,
  title    = "On the Relative Importance of Visual and Spatial Audio Rendering
              on {VR} Immersion",
  author   = "Potter, Thomas and Cvetković, Zoran and De Sena, Enzo",
  journal  = "Frontiers in Signal Processing",
  volume   =  2,
  year     =  2022
}

@ARTICLE{Pulopulos2020-ls,
  title    = "Cortisol response to stress: The role of expectancy and
              anticipatory stress regulation",
  author   = "Pulopulos, Matias M and Baeken, Chris and De Raedt, Rudi",
  journal  = "Horm. Behav.",
  volume   =  117,
  pages    =  104587,
  month    =  jan,
  year     =  2020,
  language = "en"
}

@BOOK{Zotter2019-ub,
  title     = "Ambisonics",
  author    = "Zotter, Franz and Frank, Matthias",
  publisher = "Springer International Publishing",
  month     =  may,
  year      =  2019
}

@INPROCEEDINGS{Duh2006-zo,
  title     = "Usability evaluation for mobile device: a comparison of
               laboratory and field tests",
  author    = "Duh, Henry Been-Lirn and Tan, Gerald C B and Chen, Vivian
               Hsueh-Hua",
  booktitle = "Proceedings of the 8th conference on Human-computer interaction
               with mobile devices and services",
  publisher = "Association for Computing Machinery",
  address   = "New York, NY, USA",
  pages     = "181--186",
  series    = "MobileHCI '06",
  month     =  sep,
  year      =  2006
}

@ARTICLE{Slater1993-mf,
  title     = "Representations systems, perceptual position, and presence in
               immersive virtual environments",
  author    = "Slater, Mel and Usoh, Martin",
  journal   = "Presence",
  publisher = "MIT Press - Journals",
  volume    =  2,
  number    =  3,
  pages     = "221--233",
  month     =  jan,
  year      =  1993,
  language  = "en"
}

@ARTICLE{Insko2003-zh,
  title     = "Measuring presence: Subjective, behavioral and physiological
               methods",
  author    = "Insko, Brent E",
  journal   = "Being there: Concepts, effects and measurements of user presence
               in synthetic environments.",
  publisher = "IOS Press, xx",
  address   = "Amsterdam, Netherlands",
  volume    =  321,
  pages     = "109--119",
  year      =  2003
}

@ARTICLE{Choi2005-xs,
  title    = "A catalog of biases in questionnaires",
  author   = "Choi, Bernard C K and Pak, Anita W P",
  journal  = "Prev. Chronic Dis.",
  volume   =  2,
  number   =  1,
  pages    = "A13",
  month    =  jan,
  year     =  2005,
  language = "en"
}

@ARTICLE{Olaverri2019-ic,
  title    = "The {ACFR} Centre: {ITS} Group [ITS Research Lab]",
  author   = "Olaverri, Cristina",
  journal  = "IEEE Intell. Transp. Syst. Mag.",
  volume   =  11,
  number   =  3,
  pages    = "235--240",
  year     =  2019
}

@ARTICLE{Lim2008-fl,
  title     = "The anatomy of prototypes",
  author    = "Lim, Youn-Kyung and Stolterman, Erik and Tenenberg, Josh",
  journal   = "ACM Trans. Comput. Hum. Interact.",
  publisher = "Association for Computing Machinery",
  volume    =  15,
  number    =  2,
  pages     = "1--27",
  month     =  jul,
  year      =  2008
}

@BOOK{Buxton2007-bw,
  title     = "Sketching User Experience: Getting the Design Right and the Right
               Design",
  author    = "Buxton, B",
  publisher = "Morgan Kaufmann",
  month     =  jan,
  year      =  2007
}

@INPROCEEDINGS{Buchenau2000-cu,
  title     = "Experience Prototyping",
  author    = "Buchenau, Marion and Suri, Jane Fulton",
  booktitle = "Proceedings of the Conference on Designing Interactive Systems",
  publisher = "unknown",
  pages     = "424--433",
  month     =  aug,
  year      =  2000
}

@ARTICLE{Flohr2022-hz,
  title     = "The Value of Context-Based Interface Prototyping for the
               Autonomous Vehicle Domain: A Method Overview",
  author    = "Flohr, Lukas A and Wallach, Dieter P",
  journal   = "Multimodal Technologies and Interaction",
  publisher = "Multidisciplinary Digital Publishing Institute",
  volume    =  7,
  number    =  1,
  pages     =  4,
  month     =  dec,
  year      =  2022,
  language  = "en"
}

@INPROCEEDINGS{Chamberlain2012-vb,
  title     = "Research in the wild: understanding 'in the wild' approaches to
               design and development",
  author    = "Chamberlain, Alan and Crabtree, Andy and Rodden, Tom and Rogers,
               Yvonne",
  booktitle = "Proceedings of the Designing Interactive Systems Conference",
  publisher = "unknown",
  pages     = "795--796",
  month     =  jun,
  year      =  2012
}

@INPROCEEDINGS{Paulos2005-em,
  title     = "Urban probes",
  author    = "Paulos, Eric and Jenkins, Tom",
  booktitle = "Proceedings of the SIGCHI Conference on Human Factors in
               Computing Systems",
  publisher = "ACM",
  address   = "New York, NY, USA",
  month     =  apr,
  year      =  2005,
  language  = "en"
}

@INPROCEEDINGS{Yeo2020-zm,
  title     = "Toward Immersive Self-Driving Simulations: Reports from a User
               Study across Six Platforms",
  author    = "Yeo, Dohyeon and Kim, Gwangbin and Kim, Seungjun",
  booktitle = "Proceedings of the 2020 CHI Conference on Human Factors in
               Computing Systems",
  publisher = "Association for Computing Machinery",
  address   = "New York, NY, USA",
  pages     = "1--12",
  series    = "CHI '20",
  month     =  apr,
  year      =  2020
}

@INPROCEEDINGS{Hoggenmuller2014-dd,
  title     = "{LightSet}: enabling urban prototyping of interactive media
               façades",
  author    = "Hoggenmüller, Marius and Wiethoff, Alexander",
  booktitle = "Proceedings of the 2014 conference on Designing interactive
               systems",
  publisher = "Association for Computing Machinery",
  address   = "New York, NY, USA",
  pages     = "925--934",
  series    = "DIS '14",
  month     =  jun,
  year      =  2014
}

@ARTICLE{Tran2023-qv,
  title     = "Simulating Wearable Urban Augmented Reality Experiences in {VR}:
               Lessons Learnt from Designing Two Future Urban Interfaces",
  author    = "Tran, Tram Thi Minh and Parker, Callum and Hoggenmüller, Marius
               and Hespanhol, Luke and Tomitsch, Martin",
  journal   = "Multimodal Technologies and Interaction",
  publisher = "Multidisciplinary Digital Publishing Institute",
  volume    =  7,
  number    =  2,
  pages     =  21,
  month     =  feb,
  year      =  2023,
  language  = "en"
}

@INPROCEEDINGS{M_Faas2021-ox,
  title     = "Calibrating Pedestrians' Trust in Automated Vehicles: Does an
               Intent Display in an External {HMI} Support Trust Calibration and
               Safe Crossing Behavior?",
  author    = "M. Faas, Stefanie and Kraus, Johannes and Schoenhals, Alexander
               and Baumann, Martin",
  booktitle = "Proceedings of the 2021 CHI Conference on Human Factors in
               Computing Systems",
  publisher = "Association for Computing Machinery",
  address   = "New York, NY, USA",
  number    = "Article 157",
  pages     = "1--17",
  series    = "CHI '21",
  month     =  may,
  year      =  2021
}

@INPROCEEDINGS{Faas2020-gx,
  title     = "A Longitudinal Video Study on Communicating Status and Intent for
               Self-Driving Vehicle – Pedestrian Interaction",
  author    = "Faas, Stefanie M and Kao, Andrea C and Baumann, Martin",
  booktitle = "Proceedings of the 2020 CHI Conference on Human Factors in
               Computing Systems",
  publisher = "Association for Computing Machinery",
  address   = "New York, NY, USA",
  pages     = "1--14",
  series    = "CHI '20",
  month     =  apr,
  year      =  2020
}

@INPROCEEDINGS{Hollander2019-po,
  title     = "A pedestrian perspective on autonomous vehicles",
  author    = "Holländer, Kai",
  booktitle = "Proceedings of the 24th International Conference on Intelligent
               User Interfaces: Companion",
  publisher = "Association for Computing Machinery",
  address   = "New York, NY, USA",
  pages     = "149--150",
  series    = "IUI '19",
  month     =  mar,
  year      =  2019
}

@INPROCEEDINGS{Risto2017-gr,
  title     = "Human-Vehicle Interfaces: The Power of Vehicle Movement Gestures
               in Human Road User Coordination",
  author    = "Risto, Malte and Emmenegger, Colleen and Vinkhuyzen, Erik and
               Hollan, Jim",
  booktitle = "Driving Assessment Conference",
  publisher = "unknown",
  pages     = "186--192",
  month     =  nov,
  year      =  2017
}

@INPROCEEDINGS{Dey2017-tz,
  title     = "Pedestrian Interaction with Vehicles: Roles of Explicit and
               Implicit Communication",
  author    = "Dey, Debargha and Terken, Jacques",
  booktitle = "Proceedings of the 9th International Conference on Automotive
               User Interfaces and Interactive Vehicular Applications",
  publisher = "Association for Computing Machinery",
  address   = "New York, NY, USA",
  pages     = "109--113",
  series    = "AutomotiveUI '17",
  month     =  sep,
  year      =  2017
}

@INPROCEEDINGS{Pelikan2023-tb,
  title     = "Designing Robot Sound-In-Interaction: The Case of Autonomous
               Public Transport Shuttle Buses",
  author    = "Pelikan, Hannah R M and Jung, Malte F",
  booktitle = "Proceedings of the 2023 ACM/IEEE International Conference on
               Human-Robot Interaction",
  publisher = "Association for Computing Machinery",
  address   = "New York, NY, USA",
  pages     = "172--182",
  series    = "HRI '23",
  month     =  mar,
  year      =  2023
}

@INPROCEEDINGS{Moore2019-ul,
  title     = "The Case for Implicit External Human-Machine Interfaces for
               Autonomous Vehicles",
  author    = "Moore, Dylan and Currano, Rebecca and Strack, G Ella and Sirkin,
               David",
  booktitle = "Proceedings of the 11th International Conference on Automotive
               User Interfaces and Interactive Vehicular Applications",
  publisher = "Association for Computing Machinery",
  address   = "New York, NY, USA",
  pages     = "295--307",
  series    = "AutomotiveUI '19",
  month     =  sep,
  year      =  2019
}

@ARTICLE{Weir2010-kw,
  title    = "Application of a driving simulator to the development of
              in-vehicle human–machine-interfaces",
  author   = "Weir, David H",
  journal  = "IATSS Research",
  volume   =  34,
  number   =  1,
  pages    = "16--21",
  month    =  jul,
  year     =  2010
}

@INPROCEEDINGS{Patrao2015-ml,
  title     = "How to Deal with Motion Sickness in Virtual Reality",
  author    = "Patrão, Bruno and Pedro, Samuel Lago and Menezes, Paulo",
  booktitle = "Encontro Português de Computação Gráfica e Interação",
  publisher = "unknown",
  month     =  nov,
  year      =  2015
}

@ARTICLE{Skola2020-jk,
  title    = "Virtual Reality with 360-Video Storytelling in Cultural Heritage:
              Study of Presence, Engagement, and Immersion",
  author   = "Škola, Filip and Rizvić, Selma and Cozza, Marco and Barbieri,
              Loris and Bruno, Fabio and Skarlatos, Dimitrios and Liarokapis,
              Fotis",
  journal  = "Sensors",
  volume   =  20,
  number   =  20,
  month    =  oct,
  year     =  2020,
  language = "en"
}

@ARTICLE{Blair2021-mp,
  title    = "Immersive {360°} videos in health and social care education: a
              scoping review",
  author   = "Blair, Carolyn and Walsh, Colm and Best, Paul",
  journal  = "BMC Med. Educ.",
  volume   =  21,
  number   =  1,
  pages    =  590,
  month    =  nov,
  year     =  2021,
  language = "en"
}

@INCOLLECTION{Jones2022-qn,
  title     = "Creating {360°} imagery",
  author    = "Jones, Phil and Osborne, Tess and Sullivan-Drage, Calla and Keen,
               Natasha and Gadsby, Eleanor",
  booktitle = "Virtual Reality Methods",
  publisher = "Bristol University Press",
  edition   =  1,
  pages     = "98--116",
  series    = "A Guide for Researchers in the Social Sciences and Humanities",
  year      =  2022
}

@ARTICLE{Sutcliffe2018-ja,
  title     = "Reflecting on the Design Process for Virtual Reality Applications",
  author    = "Sutcliffe, A G and Poullis, C and Gregoriades, A and Katsouri, I
               and Herakleous, K",
  journal   = "Int. J. Hum. Comput. Interact.",
  publisher = "Taylor \& Francis Group",
  volume    =  35,
  number    =  2,
  pages     = "1--12",
  month     =  mar,
  year      =  2018
}

@INPROCEEDINGS{Wang2022-zb,
  title     = "Pedestrian-Vehicle Interaction in Shared Space: Insights for
               Autonomous Vehicles",
  author    = "Wang, Yiyuan and Hespanhol, Luke and Worrall, Stewart and
               Tomitsch, Martin",
  booktitle = "AutomotiveUI '22: 14th International Conference on Automotive
               User Interfaces and Interactive Vehicular Applications",
  publisher = "unknown",
  pages     = "330--339",
  month     =  sep,
  year      =  2022
}

@INPROCEEDINGS{Sutcliffe2005-lq,
  title     = "Comparing Interaction in the Real World and {CAVE} Virtual
               Environments",
  author    = "Sutcliffe, Alistair and de Bruijn, Oscar and Gault, Brian and
               Fernando, Terrence and Tan, Kevin",
  booktitle = "People and Computers XVIII — Design for Life",
  publisher = "Springer London",
  pages     = "347--361",
  year      =  2005
}

@ARTICLE{Kaur2018-mj,
  title    = "Trust in driverless cars: Investigating key factors influencing
              the adoption of driverless cars",
  author   = "Kaur, Kanwaldeep and Rampersad, Giselle",
  journal  = "Journal of Engineering and Technology Management",
  volume   =  48,
  pages    = "87--96",
  year     =  2018
}

@INPROCEEDINGS{Voit2019-yp,
  title     = "Online, {VR}, {AR}, Lab, and In-Situ: Comparison of Research
               Methods to Evaluate Smart Artifacts",
  author    = "Voit, Alexandra and Mayer, Sven and Schwind, Valentin and Henze,
               Niels",
  booktitle = "Proceedings of the 2019 CHI Conference on Human Factors in
               Computing Systems",
  publisher = "Association for Computing Machinery",
  address   = "New York, NY, USA",
  pages     = "1–12",
  series    = "CHI '19",
  year      =  2019
}

@ARTICLE{Dey2020-fx,
  title    = "Taming the {eHMI} jungle: A classification taxonomy to guide,
              compare, and assess the design principles of automated vehicles'
              external human-machine interfaces",
  author   = "Dey, Debargha and Habibovic, Azra and Löcken, Andreas and
              Wintersberger, Philipp and Pfleging, Bastian and Riener, Andreas
              and Martens, Marieke and Terken, Jacques",
  journal  = "Transportation Research Interdisciplinary Perspectives",
  volume   =  7,
  pages    =  100174,
  year     =  2020
}

@ARTICLE{Hoggenmueller2022-fw,
  title     = "Designing interactions with shared {AVs} in complex urban
               mobility scenarios",
  author    = "Hoggenmueller, Marius and Tomitsch, Martin and Worrall, Stewart",
  journal   = "Front. Comput. Sci.",
  publisher = "Frontiers Media SA",
  volume    =  4,
  pages     =  866258,
  month     =  may,
  year      =  2022,
  language  = "en"
}

\end{document}